%% file: man.tex
\documentclass[a4paper,fleqn,usenatbib,useAMS]{mnras}

\usepackage{graphicx}	
\usepackage{amsmath}	
\usepackage{amssymb}	
\usepackage{multicol}        
\usepackage{bm}		
\usepackage{pdflscape}	
\usepackage{float}
\usepackage{longtable}
\usepackage{color}
\usepackage{geometry}
\usepackage{deluxetable}
\usepackage{subcaption}
\usepackage{comment}

\newcommand{\wCen}{$\omega$~Cen}
\newcommand{\wcen}{$\omega$~Cen}
\newcommand{\wCent}{$\omega$~Centauri}
\newcommand{\amin}{$^{\prime}$}

\newcommand{\secspt}{$\buildrel{\prime\prime}\over .$}

\newcommand{\hours}{$^h$}
\newcommand{\minutes}{$^m$}
\newcommand{\secondspt}{$\buildrel{s}\over .$}
\newcommand{\lx}{L$_x$}
\newcommand{\fx}{${f}_X$}
\newcommand{\x}{$\times$}
\newcommand{\about}{$\sim$}
\newcommand{\ergs}{erg s$^{-1}$}
\newcommand{\ergscmsq}{erg s$^{-1}$ cm$^{-2}$}
\newcommand{\ergscmsqs}{ergs cm$^{-2}$ s$^{-1}$}
\newcommand{\rc}{$r_c$}
\newcommand{\rh}{$r_h$}
\newcommand{\ha}{$H\alpha$}
\newcommand{\msun}{M$_\odot$}
\newcommand{\simless}{$\la$}
\newcommand{\simgreat}{$\ga$}
\newcommand{\nH}{n$_H$}
\newcommand{\degrees}{$^{\rm o}$}

\usepackage[T1]{fontenc}
\usepackage{ae,aecompl}

\usepackage{newtxtext,newtxmath}

\title[Deep \textit{Chandra} Survey of $\omega$ Cen]{A Deep X-ray Survey of the Globular Cluster Omega Centauri}

\author[S.\ Henleywillis et.\ al]{Simon Henleywillis,$^{1}$\thanks{Contact e-mail: \href{mailto:s.henleywillis@mars.ucc.ie}{s.henleywillis@mars.ucc.ie}} 
                                    Adrienne M.\ Cool,$^{2}$\thanks{Contact e-mail: \href{mailto:cool@sfsu.edu}{cool@sfsu.edu}}
                                    Daryl Haggard,$^{3}$
                                    Craig Heinke,$^{4}$
                                    \newauthor Paul Callanan,$^{1}$
                                    and Yue Zhao$^{4}$
\\
$^{1}$Department of Physics, University College Cork, College Road, Cork, Ireland\\
$^{2}$Department of Physics and Astronomy, San Francisco State University, 1600 Holloway Avenue, San Francisco, CA 94132, USA\\
$^{3}$McGill Space Institute, 3550 University Street, Montreal, QC H3A 2A7, Canada\\
$^{4}$Department of Physics, University of Alberta, CCIS 4-183, Edmonton, AB T6G 2E1, Canada
}

\date{Last updated 2015 May 22; in original form 2013 September 5}

\pubyear{2018}

\begin{document}
\label{firstpage}
\pagerange{\pageref{firstpage}--\pageref{lastpage}}
\maketitle

\begin{abstract}

We identify 233 X-ray sources, of which 95 are new, in a 222~ks
exposure of Omega Centauri with the \textit{Chandra X-ray Observatory}'s ACIS-I detector.  The
limiting unabsorbed flux in the core is \fx (0.5--6.0~keV) $\simeq$
3\x 10$^{-16}$~\ergscmsq\ (\lx\ $\simeq$ 1\x 10$^{30}$~\ergs\ at 5.2~kpc).
We estimate that \about $60\pm 20$ of these are cluster members, of
which \about 30 lie within the core (\rc\ $=$ 155~arcsec), and another
\about 30 between 1--2 core radii.  We identify four new optical
counterparts, for a total of 45 likely identifications.  Probable
cluster members include 18 cataclysmic variables (CVs) and CV
candidates, one quiescent low-mass X-ray binary, four variable stars,
and five stars that are either associated with \wcen's anomalous red
giant branch, or are sub-subgiants.  We estimate that the cluster
contains $40\pm 10$ CVs with \lx\ $>$ 10$^{31}$~\ergs, confirming
that CVs are underabundant in \wcen\ relative to the field.  Intrinsic
absorption is required to fit X-ray spectra of six of the nine
brightest CVs, suggesting magnetic CVs, or high-inclination systems.
Though no radio millisecond pulsars (MSPs) are currently known in \wcen, more than 30 unidentified
sources have luminosities and X-ray colours
like those of MSPs found in other globular
clusters; these could be responsible for the \textit{Fermi}-detected gamma-ray emission from
the cluster.
Finally, we identify a CH star as the counterpart to the
second-brightest X-ray source in the cluster and argue that it is a
symbiotic star.  This is the first such giant/white dwarf binary to be
identified in a globular cluster.

\end{abstract}




\section{Introduction}

Significant progress has been made in recent years in unraveling the
complex interplay between stellar dynamics and stellar evolution in
globular clusters \citep[e.g.][]{Wang2016, Rodriguez2016}.  On the
observational side, critical insights have come from X-ray imaging,
which reveals many of the binary stars that drive cluster evolution at
late times.  The nearest globular clusters are prime targets for such
studies.  For these clusters, long exposures with the \textit{Chandra X-ray
Observatory} can sample luminosities as faint as \lx\ \about\ $10^{30}$~\ergs, which
enables compilation of near-complete samples of compact binaries
\citep[e.g.][]{Grindlay2001, Pooley2003, Heinke2005}.  \textit{Chandra} also
brings within reach significant numbers of main-sequence and/or
subgiant binaries that reveal themselves through enhanced coronal
activity \citep[e.g.][]{Bassa2004, Cohn2010}.

Among nearby clusters, \wcen\ stands out.  It is the most luminous of
our Galaxy's globular clusters \citep[M$_V$ $= -10.26$;][]{Harris2010}
and, with a mass of \about $3\times 10^6$~\msun\ \citep{DSouza2013,Baumgardt2017},
is the second most massive globular cluster in the entire Local Group.
Well-fit by a King-Michie model \citep{Trager1995, Ferraro2006}, it
has an enormous core with radius \rc\ $\simeq$ 155 arcsec\ $=$ 3.9 pc at an
 assumed distance of 5.2 kpc \citep{Trager1995, Harris2010}.  As a
result, a high rate of stellar interactions is predicted
\citep{Verbunt1988, DiStefano1994, Davies1995}, despite the cluster's relatively
modest central density \citep[$\rho _0$ \about\ 3\x $10^3$
  \msun/pc$^3$;][]{Pryor1993}.

Omega Cen was also the first cluster to show signs of multiple stellar
populations \citep{Freeman1975, Norris1995, Anderson2002}, a
phenomenon now well-documented in nearly every closely-examined
globular cluster in the Milky Way \citep{Piotto2015}.  In \wcen's
case, the spread in chemical composition is unusually large, and
includes widely disparate Fe abundances among its multiple populations
\citep[see review by][]{Gratton2004}.  These curious properties have
led to the suggestion that \wcen\ may not be a globular cluster at
all, but instead the remnant nucleus of a dwarf galaxy captured by the
Milky Way \citep{Norris1996, Lee1999, Bekki2003}.

Observations with successive generations of X-ray observatories have
revealed increasingly large numbers of sources in and toward \wcen.
Five were found with \textit{Einstein/IPC} \citep{Hertz1983}, 21 with
\textit{ROSAT/PSPC} \citep{Johnston1994, Verbunt2000}, and 146 with \textit{XMM-Newton}
\citep{Gendre2003}.  In each case, roughly one third to one-half of
the sources lay within about 3 core radii of the cluster centre, where
membership in the cluster becomes more probable.  The first \textit{Chandra}
observation of \wcen, a \about 70~ksec exposure with the Advanced CCD
Imaging Spectrometer's imaging array (ACIS-I), expanded the number of X-ray sources to
180, \about 150 of which lie interior to 3\rc\ \citep[][see also \citealt{Cool2002}]{Haggard2009}.

The ability to observe faint stars even in the cluster core with
the \textit{Hubble Space Telescope (HST)} has made it possible to identify optical
counterparts for a significant number of these sources
\citep{Carson2000, Haggard2004, Cool2013}.  Optical counterparts are
essential for classifying the objects that are responsible for the
X-ray emission; very few objects can be uniquely identified on the
basis of X-rays alone, quiescent low-mass X-ray binaries being a rare
exception \citep{Rutledge2002}.  Innovations in ground-based
photometric techniques have also uncovered a large number of variable
stars in \wcen\ \citep{Kaluzny1996, Kaluzny2004}, some of which are
detectable in X-rays at the sensitivity levels achieved by \textit{XMM-Newton} and
\textit{Chandra}.

In 2012 we obtained a second epoch of \textit{Chandra} imaging of \wcen.  A
primary goal of the study, which had a factor of about three more
exposure time than the original study, was to search for -- or put
limits on -- X-ray emission from a potential intermediate-mass black
hole (IMBH).  Focusing on the central part of the ACIS-I field of
view, we found no X-ray emission coincident with the cluster centre,
and concluded that if an IMBH does exist in \wcen\ then it must be
experiencing very little or very inefficient accretion
\citep{Haggard2013}.

Here we report our analysis of the full 2012 ACIS-I \wcen\ dataset.
In Section 2 we describe the observations, how we processed the data,
and the methods by which we identified sources and measured their
properties.  We compare the results to those obtained from the first
epoch of observations in Section 3.  In Section 4 we compare the
observed sources to predicted numbers of active galactic nuclei (AGN)
and X-ray-detectable foreground stars both to determine how many sources
are cluster members and to assess their radial and luminosity
distribution.  In Section 5 we summarise existing optical
identifications and add several more by comparing the new source list
to existing catalogues of variable stars.  A particularly interesting
new optical ID, the first symbiotic star found in a globular cluster,
is described in Section 6.  In Section 7 we present an X-ray
colour--magnitude diagram along with spectral and timing analyses of
several of the brightest sources in the cluster (additional spectral
and timing analyses will appear elsewhere).  We discuss the findings in
Section 8 and summarise them in Section 9.

\section{X-RAY OBSERVATIONS AND SOURCE LIST}
\label{sec:xray_obs}

We obtained two long exposures of \wCen\ using the imaging array of
the \textit{Chandra X-ray Observatory}'s Advanced CCD Imaging
Spectrometer (ACIS-I) on 2012 April 16$-$17. The datasets
have a combined exposure time of $\sim$222 ksec
(173.7 and 48.5 ksec for ObsIDs 13726 and 13727, respectively).  The
cluster centre was placed near the middle of the imaging array's
$\sim$16.9 \x\ 16.9 arcmin field of view (see Fig.~1), which is
comprised of a 2$\times$2 array of 1024$\times$1024 chips with
$\sim$0.49 arcsec pixels.  To optimise the detection of faint sources,
we used the `very faint' telemetry format in the Timed Exposure
mode, which permits improved screening for particle events in
post-processing\footnote{See
  http://cxc.cfa.harvard.edu/proposer/POG/html/chap6.html}.
  
  \begin{figure}
\centering
  \includegraphics[width=\linewidth]{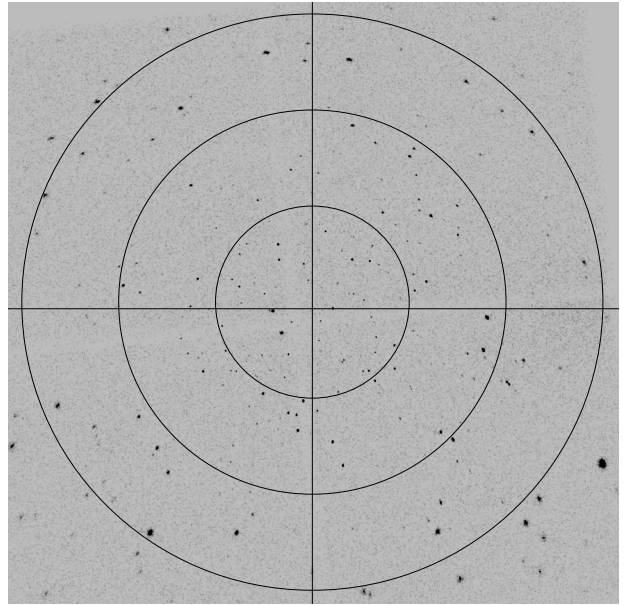}
  \caption{Smoothed 222 ksec \textit{Chandra} ACIS-I image of \wcen.  Circles indicate 1, 2, and 3 core radii (\rc $=$ 155 arcsec), and the crosshair indicates the cluster centre. The cluster's half-mass radius is 5.0 arcmin, i.e.\ just under 2 core radii.}
  \label{fig:ACIS_3rc}
\end{figure}

We processed the two raw datasets using CIAO's
\textsc{chandra$\_$repro}.  This tool filters events by grade and
status, selects good time intervals, and generates a bad pixel file.
Reprocessing also enabled us to take advantage of the `very faint' mode and
flag events associated with cosmic rays, which significantly reduced
the background rate in the final level$=$2 event files.  In preparation
for source detection, we combined the two level$=$2 events files using
\textsc{merge$\_$obs} to create an unbinned merged image in the
0.5--4.5~keV band for the full dataset.  Eliminating counts above 4.5~keV
at this stage further reduced the background with relatively
little effect on the counts associated with most sources.  This tool
also generated an exposure map suitable for use with
\textsc{wavdetect} (see below).  We generated a point spread
function (PSF) map at 1.9 keV for an encircled counts fraction (ECF)
of 0.5.  Since PSF maps cannot be generated directly for
merged images, we first created separate PSF maps for each ObsID and
then used \textsc{dmimgcalc} to produce an exposure-weighted sum of
the two.  The choice of ECF$=$0.5 was made after
experimenting with \textsc{wavdetect} and finding that the ability to
distinguish closely-spaced on-axis sources was sensitive to this
parameter.  

We used CIAO's \textsc{wavdetect} tool to search for sources in the
0.5--4.5 keV merged image, supplying it with the exposure map and PSF
map described above.  After experimenting with various combinations of
parameters, we adopted wavelet scales \texttt{[1, 2, 4, 8]} and a
source significance threshold of \texttt{10}$^{\texttt{-6}}$.  This set
of parameters was effective at picking up close pairs inside the
half-mass radius while limiting spurious off-axis sources.

Table~3 lists the full set of 233 sources that appear in the 2012 data
set, 95 of which are new.  We adopt the same labelling convention used
by \citet{Haggard2009}, who identified 180 sources in the first epoch
of \textit{Chandra} data, 138 of which are recovered here (see Section 3).  For new sources, we
assign a three-character ID based on radial offset from the centre of
the cluster\footnote{To preserve consistency with \citet{Haggard2009},
  we assigned source IDs using the same cluster centre adopted in that
  paper (R.A. $=$ 13\hours26\minutes45\secondspt89, Dec. $=$ $-$47\degrees28\amin36\secspt7).} and azimuthal angle (see column 1 of Table 3).  The first
character is a number between 1 and 10 equal to the radial offset rounded
to 1 arcmin.  The second character is a number between 1 and 4
identifying the quadrant in which the source falls.  The third
character is a letter based on azimuthal angle, increasing
counterclockwise.  Original labels were kept for re-detected sources;
new source names begin with the next available letter and are shown in
bold face.  Coordinates for the sources, as reported by
\textsc{wavdetect}, are given in column 2 of Table 3.  Uncertainties
on these positions, at 95 per cent confidence, computed from
\textsc{wavdetect} source counts and off-axis angle following the
empirical formula developed by \citet[][see their Eq.~5]{Hong2005}, are
listed in column 3.  Radial offsets from the cluster centre in units
of the core radius ($r_c=$ 155 arcsec) are listed in column 4.  For these
offsets we adopted the centre measured by \citet{Anderson2010}:
R.A. $=$ 13\hours26\minutes47\secondspt24, Dec. $=$ $-$47\degrees28\amin46\secspt45.

To determine source counts and fluxes we used the CIAO tool
\textsc{srcflux}.  We extracted counts in apertures that increase in
size with increasing distance from the optical axis to ensure
encircled energies $\ge$ 50 per cent at 1.9 keV over the full field (and
$\ge$ 80 per cent for most sources; see Table 1).  For purposes of
comparison to \citet{Haggard2009} and to long-exposure \textit{Chandra} studies
of other nearby globular clusters \citep[e.g.][]{Heinke2005,
  Bogdanov2010}, we extracted counts in several bands: `medium'
(0.5--4.5~keV), `soft' (0.5--1.5~keV), `hard' (1.5--6.0~keV),
`wide' (0.5--6.0~keV), as well as in the 0.5--2.0~keV soft band
used for the \textit{Chandra} Deep Field South \citep[CDF-S;][]{Luo2008}.  The
first numbers listed in the columns 5--7 of Table 3 are the merged raw
counts in the first three of these bands.  We note that
\textsc{srcflux} does not operate on merged images; we therefore
extracted raw source counts from the two level$=$2 events files
individually and summed the results.

\input table2_FINAL.tex

The second entries in columns 5--7 incorporate three adjustments to
the raw counts: background subtraction, aperture correction, and an
exposure correction.  We measured background values for each source
separately in annuli centred on the sources and sufficiently large to
yield background values accurate to $\sim$ 10 per cent and sufficiently far
from the sources that no more that 3 per cent of source counts fall in the
annulus (and much less than 1 per cent in most cases; see Table 1).
Several background annuli encompass a neighboring source; in these
cases we excluded the imposing source's region from the annulus before
measuring the background level.  A small number of faint sources for
which background subtraction produced negative values in the soft or
hard bands are listed in Table 3 as having zero corrected counts.
Following background subtraction, we made aperture corrections by
dividing the background-subtracted counts by the 1.9 keV encircled
energy associated with each individual source, as reported by
\textsc{srcflux}.  Finally, to correct for differing effective areas
across the field (e.g.\ due to chip gaps), we normalised all
sources to the typical on-axis value of 350 cm$^2$.  Corrected counts
are thus given by:

\begin{equation}
\text{corrected counts} = \left(\frac{\texttt{NET\_COUNTS}}{\texttt{PSFFRAC}}\right)\left(\frac{350\text{ cm}^2}{\texttt{MEAN\_SRC\_EXP}}\right)\label{eq:cnts-corr}
\end{equation}

The log of the ratio of the corrected counts in the soft vs. hard bands
is given in Table 3, column 8.

To convert corrected counts to unabsorbed fluxes we chose a
power law model with a photon index of 1.4, appropriate for the large
number of AGN expected to be included among the sources \citep[see
  Section~4;][]{Giacconi2001}.  We assumed a hydrogen column toward
\wCen\ of n$_H$ $=$ 9$\times$10$^{20}~$cm$^{-2}$.  The latter was
derived assuming \textit{E(B$-$V)} $=$ 0.11 \citep{Lub2002}, \textit{A$_V$/E(B$-$V)} $=$
3.1 \citep{Cardelli1989}, and n$_H$/\textit{A$_V$} $=$ $2.81\times$10$^{21}$
\citep{Bahramian2015}.  We used a Tuebingen-Boulder interstellar
medium absorption model and abundances from \citet{Wilms2000}
\citep[see][]{Heinke2014}, which are incorporated into
\textsc{srcflux} as \texttt{tbabs} and \texttt{wilm}, respectively.
The resulting unabsorbed fluxes for the medium, wide, and CDF-S soft bands
are reported in Table 3 columns 9--11, respectively.  These fluxes are
exposure-time weighted averages of fluxes determined by running
\textsc{srcflux} on the two level$=$2 events files separately.

In Fig.~2 we plot the 0.5--4.5 keV fluxes from Table 3 for the 233
sources reported here against their radial offsets from the cluster
centre, 20 arcsec from the ACIS-I aimpoint.  The 138
recovered sources are shown as blue circles and the 95 new sources
are shown as red triangles.  Fluxes measured in the first-epoch data
for 42 sources that were not recovered in the new data are shown as
black dots.  These bring the total number of X-ray sources known in
and toward \wCen\ to 275.  The impact of the off-axis broadening of
the PSF is clearly visible: while the limiting flux reaches
\fx (0.5--4.5~keV) \about\ 2\x $10^{-16}$~\ergs\ near the cluster centre, the limit
increases steadily with radius, reaching \fx (0.5--4.5 keV) \about\ 2\x $10^{-15}$
\ergs\ near the edge of the ACIS-I field.

\begin{figure}
\centering
  \includegraphics[width=\linewidth]{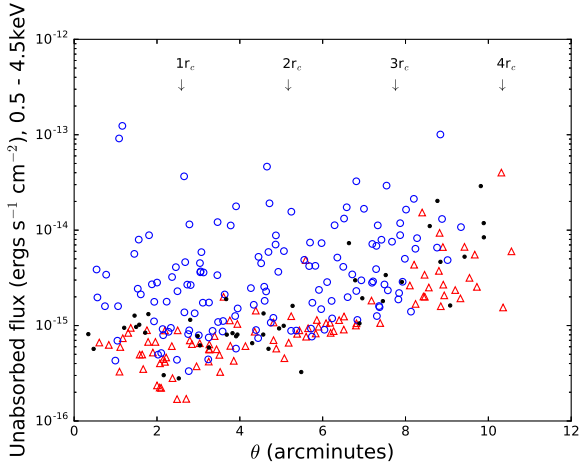}
\caption{X-ray flux as a function of angular distance from
  the cluster centre for 275 known \textit{Chandra} sources in and toward \wcen.
  Red triangles represent 95 sources newly found in this study.  Blue
  circles are 138 sources reported by \citet{Haggard2009} and recovered in
  the present study.  Black dots are 42 sources found by \citet{Haggard2009}
  that were not detected in the 2012 dataset, including six that lie 
  outside the ACIS-I field of view of the new data.  Fluxes for these
  sources were remeasured  from the 2000 dataset using methods identical 
  to those used for the 2012 data.}
  \label{fig:f_c_g}
\end{figure}

\section{COMPARISON WITH FIRST-EPOCH OBSERVATIONS}

The field of view of the present observations overlaps with the
first-epoch observations by approximately 88 per cent.  In principle the
two could be combined to yield a somewhat deeper image; however,
given the significant change in sensitivity of ACIS-I in the $\sim$
12~yr time period separating the two observations and the strong
wavelength dependence of those changes,\footnote{See
  http://cxc.harvard.edu/ciao/why/acisqecontam.html for details.}
characterization of the resulting sources would be problematic.
Instead we compare results obtained by reducing the two epochs
separately.

To ensure that we are making valid comparisons, we have re-extracted
source counts and recomputed fluxes from the 2000 dataset using
procedures identical to those used for the 2012 data, beginning with
the \citet{Haggard2009} source list.  This is important given that the
details of the analysis methods adopted here differ somewhat from
those used by \citet{Haggard2009}.  In Fig.~3 we plot the 2012
0.5--6.0 keV fluxes as a function of the 2000 fluxes in the same band for
sources that appear in both datasets.  Ratios of these fluxes are also
listed in column 12 of Table 3.  Here it can be seen that while many
sources have 2012 fluxes that are consistent with their values 12
years earlier (solid red line), a significant number have changed in
brightness by a factor of 2--3. The median flux ratio for the full
set of 138 sources is 0.94. A small number of sources show
variations in excess of 3 and up to a factor of \about 10.  The source
with the largest change in flux is a known cataclysmic variable (CV;
ID $=$ 12a); its flux dropped by a factor of \about 15 from 2000 to
2012.

\begin{figure}
\centering
  \includegraphics[width=\linewidth]{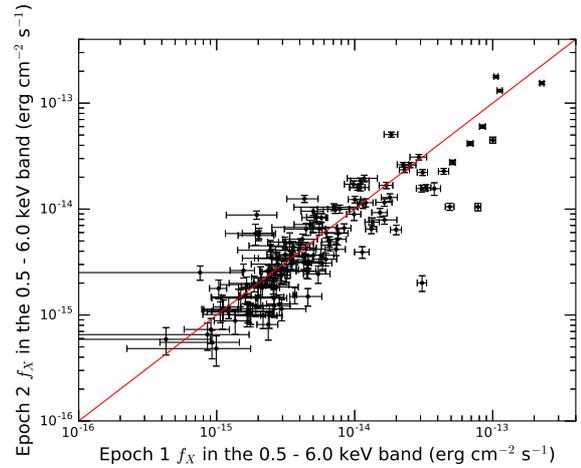}
\caption{Unabsorbed wide-band 2012 fluxes vs.\ 2000 fluxes for the 138
  sources that the two observations have in common.  Fluxes have been
  derived assuming a power law with photon index $\Gamma = 1.4$ (see
  Section 3). The red line denotes equality.  Error bars represent 
  statistical uncertainties only and do not account for systematic errors 
  associated with assuming a single spectral shape for all sources.}
  \label{fig:f_sq}
\end{figure}

The assumption of an identical spectrum for all sources in computing
the fluxes (see Section 2) means that not all of the apparent flux
changes seen in Fig.~3 will necessarily be indicative of actual source
variability.  Sources whose spectra differ significantly from the
assumed $\Gamma = 1.4$ power-law spectrum will have different inferred
fluxes in the absence of any real change in brightness owing to the
change in sensitivity of the instrument as a function of wavelength in
the 12~yr interval between the two sets of observations.
Nevertheless, the fact that the median flux ratio for the full set of
138 sources is close to unity suggests that the spectral model we have
adopted is reasonable for the bulk of the sources.

Further evidence for considerable flux variations for the sources in
this field comes from the fact that 36, i.e.\ \about 20 per cent, of the
sources detected in 2000 are not detected in the 2012 data despite
lying within the 2012 ACIS-I FOV (see Fig.~2).  This is in spite of
the fact that the exposure time for the 2012 observations is more than
a factor of three longer than for the first-epoch observations
(although the loss of sensitivity offsets the increased exposure time
somewhat).  Conversely, 82 of the sources reported here are in the
earlier ACIS-I FOV but were not detected in that observation.  While
most of the latter are faint sources that would have been below the
flux limit in the shorter exposures, others are clearly bright enough
that they should have been detected in the absence of intrinsic
variation (see Fig.~2).  

Variability in X-rays is characteristic of nearly all types of objects
we expect to detect in these data, including cataclysmic variables and
active binaries in the cluster, AGN in the background, and X-ray
bright stars in the foreground; we analyse several of the brightest
sources in Section 7 below.  An exception is the quiescent neutron
star (source 44e) in \wcen\ \citep{Rutledge2002, Gendre2003,
  Haggard2004}.  If this object is truly quiescent and is not
accreting, then the thermal X-ray emission from the hot neutron star's
surface should be unchanged from 2000 to 2012.  The analysis of the
spectrum and flux of this source by \citet{Heinke2014}, which compared
the two \textit{Chandra} epochs and the \textit{XMM-Newton} epoch, shows no detectable
variation, consistent with this picture.

\section{X-RAY SOURCE MEMBERSHIP}

Given the length of the ACIS-I exposures and the sensitivity of the
camera, significant numbers of AGN will be present in the data, as
well as foreground stars.  The cumulative number of sources as a
function of flux in the entire 2012 field of view (FOV) is shown in Fig.~4, together
with results obtained by \citet{Luo2008} for the \textit{Chandra} Deep Field
South (CDF-S).\footnote{A table of values corresponding to the 2~Ms
  CDF-S measurements, corrected for incompleteness as shown in
  Fig.~15a of \citet{Luo2008}, was kindly provided by B.\ Luo.}  These
authors analysed a \about 2 Ms exposure of the CDF-S, detecting 578
sources in a \about 436 square arcminute field of view with limiting
sensitivity of \fx $\simeq1.9\times10^{-17}$~\ergscmsqs
in the 0.5--2.0 keV band.  Scaling the CDF-S results to the \about
286 square arcminute field of view of the present study reveals a
clear excess of sources toward \wcen\ with fluxes in excess of
\fx(0.5--2.0~keV) \about\ 2\x $10^{-16}$~\ergscmsq\ (see Fig.~4).  At
\fx(0.5--2.0~keV) $=$ 5\x $10^{-15}$~\ergscmsq, for example, the cumulative excess is a factor
of two, with 29 sources observed and only \about 14 AGN predicted.  At
\fx(0.5--2.0~keV) $=$ 5\x $10^{-16}$~\ergscmsq, the cumulative excess is still
\about 70 per cent, with 169 sources observed and only \about 100 AGN
predicted.

To examine the radial dependence of the X-ray source distribution in
\wcen\ we have divided the ACIS-I field of view into four regions: the
cluster core, two concentric annuli (1--2\rc, 2--3\rc), and a
fourth region that includes all ACIS-I area outside 3\rc\ (see
Fig.~5).  The first three regions are fully encompassed within the
ACIS-I field of view and occupy approximately 21, 63, and 105 square
arcminutes, respectively, while the fourth spans the remaining \about
97 square arcminutes.  Dividing up the field in this way enables us to
take the radial dependence of the limiting flux into account.  The
cumulative number of sources as a function of flux in the 0.5--2.0~keV
CDF-S soft band are shown as solid blue lines in Fig.~5.  The
predicted numbers of AGN \citep{Luo2008}, scaled to the area of each
region, are shown for comparison (dashed black lines) with dotted
black lines representing root N uncertainties.  The difference, i.e.\
sources not attributable to background AGN, are shown as solid red
lines.

\begin{figure}
\centering
  \includegraphics[width=0.89\linewidth]{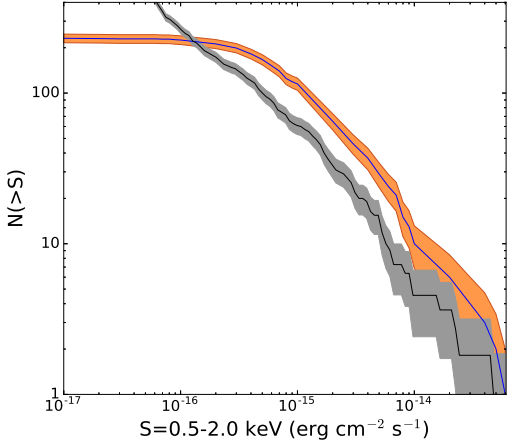}
\caption{The number of sources $N$ that are brighter than a given flux
  $S$ in the CDF-S 0.5--2.0 keV energy band (blue line with $\sqrt{N}$ error range in red).
The predicted number of AGN based on the \textit{Chandra} Deep Field South
study of \citet{Luo2008} is shown as a black line with $\sqrt{N}$ error range in grey.}
  \label{fig:agn_full}
\end{figure}

\begin{figure}
\centering
  \includegraphics[width=\linewidth]{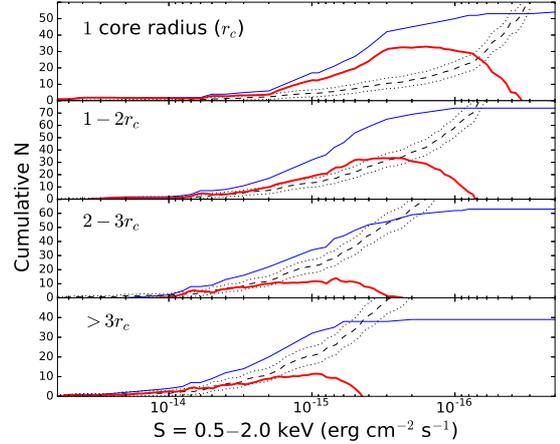}
\caption{Cumulative numbers of observed sources in four regions as a
  function of flux in the 0.5--2.0 keV band (solid blue lines).
  Expected numbers of background sources scaling from \citet{Luo2008}
  are given as dashed black lines with uncertainties presented as
  dotted black lines.  The resultant numbers of sources not attributable
  to background AGN are shown as solid red lines.}
  \label{fig:agn_radii}
\end{figure}

These plots confirm that significant numbers of X-ray sources toward
\wCen\ cannot be attributed to background AGN, and that this is true not only in
the core of the cluster, but also in the 1--2 \rc\ annulus.
Specifically, we estimate that $33\pm 4$ and $33\pm 6$ sources in these two
regions, respectively, cannot be attributed to AGN.\footnote{Taking
  account of the reduced sensitivity in chip gaps reduces the predicted
numbers of AGN in the core by 1--2 sources, and correspondingly increases
the estimated number of member sources.  Outside the core the correction is negligible.}  In the outer two
regions, the numbers are smaller but still non-negligible, with $14\pm5$,
and $12\pm 5$ sources unattributed to AGN in the 2--3~\rc\ and
$>3$\rc\ regions, respectively.  The total number of sources
unattributed to AGN is thus \about $90\pm 20$.  These numbers increase
by \about 15 sources if, like \citet{Haggard2009}, we derive AGN
estimates from \citet{Tozzi2001} rather than \citet{Luo2008}.

Estimating how many of the sources may be foreground stars is more
challenging given the variations in star counts as a function of
Galactic latitude and longitude.  A rough idea can be gleaned from the
\textit{XMM-Newton} SSC survey of the galactic plane \citep{Nebot2013}, which
combines results for fields with galactic latitudes in the range b =
12--19\degrees\ and longitudes in the range l = 54--237\degrees.  For
comparison, \wcen\ has b $=$ 15\degrees\ and l $=$ 51\degrees.
Extrapolating the \citet{Nebot2013} results for known and suspected
coronal sources at $|{\rm b}|$ = 15$^{\rm o}$ (see red dotted line in
their Fig.~16) from their \fx(0.5--2.0~keV) limit of \about 3.4 \x\ 10$^{-15}$~\ergscmsq\
to the \fx\ values at the peaks of the four red curves in
Fig.~5 suggests that perhaps 30--40 of the sources could be stars with
active coronae (\about 6, 12, 10, and 8 in the four radial regions,
respectively).  This leaves an estimated \about $54\pm 20$ sources
associated with \wcen.
This is likely to be a lower limit, considering that the extrapolation is
being made to fluxes that are as much as a factor of 10 fainter than those
sampled by \citet{Nebot2013} and will tend to overestimate the number of
coronal sources in the foreground since the line of sight to \wcen\ extends
outside of the thin disc where most foreground sources should lie.

For another perspective on source membership, we plot X-ray flux in
the CDF-S band as a function of the square of the offset from the
cluster centre in Fig.~6.  Here it can be seen that the density of
sources appears highest close to the cluster centre, dropping off with
increasing radius, whereas background and foreground sources should
spread evenly across the plot.  To investigate how the radial
distribution depends on flux, we adopt four flux bins (see dotted
lines in Fig.~6).  Three of the sources in the brightest bin [\fx(0.5--2.0~keV) $>$
3 \x\ 10$^{-14}$ \ergscmsq], which corresponds to
\lx\ \simgreat\ 10$^{32}$ \ergs\ at a distance of 5.2 kpc, are already
known to be cluster members \citep[CVs 13a and 13c, and qLMXB
  44e;][]{Carson2000,Haggard2004}.  The fourth we propose is a CH star
that is also a cluster member (see Section 6).

\begin{figure*}
\centering
  \includegraphics[width=\linewidth]{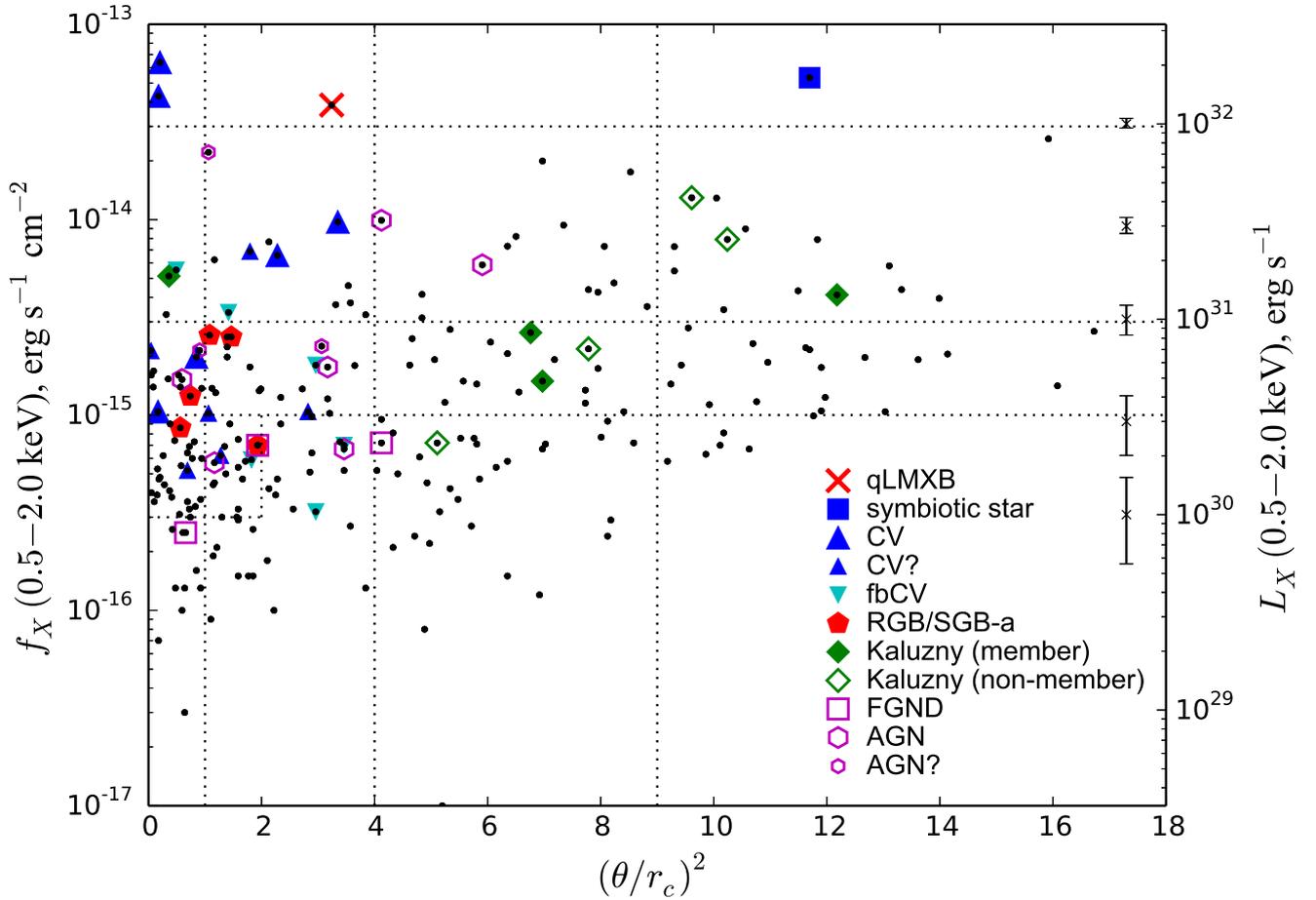}
\caption{Flux in the CDF-S 0.5--2.0 keV band as a function of the
  square of the radial offset from the cluster centre in units of the
  core radius (\rc\ $=$ 155 arcsec).  Solid coloured symbols mark
  sources for which reported optical IDs (see Figs. 5--7 in \citet{Cool2013})
  are probable cluster members;
  open symbols mark sources which are outside the cluster (based on thier optical IDs).
  Kaluzny designation refers to variable stars identified by \citet{Kaluzny1996, Kaluzny2004}.
  Representative error bars are shown at right.
  Dotted lines demarcate regions used to  analyse the radial distribution of 
  sources as a function of flux (see Section 4).}
\end{figure*}

The next bin, \fx(0.5--2.0~keV) = 3--30 \x\ 10$^{-15}$
\ergscmsq\ (\lx\ \simgreat\ 10$^{31}$--10$^{32}$ \ergs), contains 42
sources: 3, 11, 14, and 14 in the four radial bins, respectively.  In
this flux range, the census of sources should be complete out to
nearly the edge of the field (see Fig.~6).  There is no measurable
excess of sources toward the cluster centre; the relative numbers of
sources in the four bins scale roughly with the relative areas of the
bins.  However, the predicted number of AGN \citep{Luo2008} and
foreground stars \citep{Nebot2013} in the ACIS-I FOV in this flux
range are $22\pm 5$ and $7\pm 3$, respectively, which suggests that
\about $13\pm 8$ of the sources are associated with the cluster.  That
at least some are probable cluster members is corroborated by optical
IDs whose characteristics suggest membership (see solid coloured
symbols in Fig.~6).

A noticeable central concentration of sources is apparent for fainter
sources with fluxes in the bin of range \fx\ = 1--3 \x\ 10$^{-15}$ \ergscmsq,
or \lx\ \about\ 3\x 10$^{30}$ -- 1\x 10$^{31}$ \ergs\ (see Fig.~6).
In this bin, source counts are clearly incomplete in the outermost
radial bin, but should be reasonably complete inside 3\rc.  A total of
51 sources are present inside 3\rc\ in this flux range: 13 in the
core, 20 in the 1--2\rc\ annulus, and 18 in the 2--3\rc\ annulus.
Scaling from \citet{Luo2008}, \about 3, \about 8, and \about 14 AGN
are expected in these three regions, respectively.  Scaling from
\citet{Nebot2013}, another \about 1, \about 2, and \about 4 sources
are likely to be foreground stars.  While these numbers are small and
subject to considerable uncertainties, they suggest that AGN and
foreground stars could account for all or most of the sources in the
2--3\rc\ annulus.  In the core and first annulus (1--2\rc), however,
there is an excess of \about 9 and \about 10 sources, respectively.
Similar results are obtained if we simply use the sources in the
2--3\rc\ annulus as a measure of the surface density of non-members in
this flux range.  Thus we estimate that \wcen\ contains a total of
\about 20 X-ray sources with luminosities in the range
\lx\ \about\ 3\x 10$^{30}$ -- 1\x 10$^{31}$ \ergs, and that these are
roughly evenly split between the core and the 1--2\rc\ annulus.

For fluxes below \fx\ $=$ 10$^{-15}$~\ergscmsq\ (\lx\ \simless\ 3
\x\ 10$^{30}$~\ergs), source counts begin to show signs of
incompleteness even in the outer part of the 1--2\rc\ radial bin.
Still, a comparison of the number of sources in the core with fluxes
in the range \fx\ = 3\x 10$^{-16}$ -- 1\x 10$^{-15}$ \ergscmsq\ 
to the number in an annulus of the same surface area immediately
outside the core (26 and 17, respectively; see dotted regions in Fig.~6) reveals an
excess in the core.  \citet{Luo2008} predict only \about 6 AGN
in a region the size of the core in this flux range, and scaling from
\citet{Nebot2013} predicts \about 2 foreground coronal sources.  Thus
the core of \wcen\ contains \about 15--20 X-ray sources with \lx\ $\simeq$
1--3\x $10^{30}$ \ergs, and perhaps 10--15 of the \about 30
sources in the 1--2\rc\ annulus are also cluster members.

In summary, this flux-based analysis reveals a total of \about
$67\pm 20$ cluster members, consistent with the range of $54\pm 20$
derived from the Fig.~5 plots.  
This is likely to be a lower limit on the actual number of X-ray
sources in \wcen\ since the number of member sources appears to
increase steadily as the limiting flux is approached and we have
not counted sources in radial/flux bins in which incompleteness is
significant.

\section{OPTICAL IDENTIFICATIONS}

While it is clear that \wcen\ is host to a large number of X-ray
sources, optical IDs are required to determine which ones are cluster
members and what they are.  The most interesting new identification
presented here is a CH star that coincides with the second brightest
source in the ACIS-I field, 94a.  We suggest that it is a symbiotic
star -- the first such star discovered in a globular cluster -- and
discuss it further below (see Section 6).

Identifications of previously known sources given by
\citet{Haggard2009} and \citet[][see their Figs.\ 5--7]{Cool2013} are listed in column 13 of
Table 3.\footnote{Those designated `blue-only' or `\ha -only' in
  \citet{Cool2013} are least secure and are excluded here, along with
  the apparent blue straggler counterpart to 22d which has since been
  shown to be a non-member by \citet{Deveny2016}.}  These include
compact binaries (CVs and a qLMXB), AGN, foreground stars, and
variable stars both in and out of the cluster.  Subsets of these
classes whose identities are uncertain are listed with `?'
following the ID.  We also include several stars identified by
\citet{Cool2013} as possible members of \wcen's anomalously metal-rich
red-giant and/or subgiant branches (RGB/SGB-a).\footnote{One object
  (13b) which does not lie on this branch but was previously included
  in this category is excluded here.}

We searched the variable star catalogue of \citet{Kaluzny2004} for
matches to any of the 233 Cycle 13 sources. We found three new matches: 63h $=$ NV410,
73a $=$ NV369, and 91a $=$ NV379.  All are variables of unknown type with
variability periods of 1.8, 7.1, and 14.5 days, respectively.  These
are listed in column 13 of Table 3 along with five previously known matches
\citep{Haggard2009}. These new potential matches have offsets between the
optical and X-ray positions
of 1.19 arcsec, 1.46 arcsec, and 1.20 arcsec, respectively.\footnote
{Including a boresight correction does not change these values significantly.} 

The offsets between these possible new IDs and the X-ray positions
place them somewhat outside the 95 per cent error circles (see Table 3)
computed following the empirical prescription of \citet{Hong2005} -- by 
factors of 1.14, 1.97, and 1.04, respectively.  However, in view of
the large off-axis angles involved (and larger uncertainties
associated with centroiding the X-ray counts given the broad PSF),
combined with the potential for additional uncertainties associated
with the Kaluzny positions and/or proper motions of non-members, we
chose to consider matches to within 1.5 arcsec as potentially real.
No other matches of \citet{Kaluzny2004} variable stars to new \textit{Chandra}
sources were found with offsets under 2.0 arcsec.

To determine the rate of chance coincidences resulting from our choice of
1.5 arcsec error circles, we increased the error circle radii to 15 arcsec
and reran the search. This resulted in one or more Kaluzny variables landing
in 56 of the 233 expanded error circles, implying an average expected rate
of 0.56 chance coincidences with the error circles actually in use. Applying
Poisson statistics, we find that it is more likely than not that all eight
Kaluzny IDs reported here are real. However we cannot rule out the possibility
that one or two could be chance alignments (Poisson probability 32 per cent and 9 per cent,
respectively).

We also searched for matches of \textit{Chandra} sources with the catalogues of
\citet{Weldrake2007} and \citet{Lebzelter2016} and to the Henry Draper (HD) catalogue
and found no other counterparts.  Work to identify additional optical
counterparts using \textit{HST} data is underway and will be reported
elsewhere.

For some optical IDs (e.g.\ AGN), membership status is clear from
their very nature.  For others (e.g.\ CVs), a concentration toward the
cluster centre indicates that most if not all are associated with the
cluster.  In other situations cluster membership is more difficult to
assess, and in these cases proper motions are invaluable.
\citet{Bellini2009} have measured proper motions for more than 300 of
the \citet{Kaluzny2004} variable stars in the \wcen\ field, including
8 of the 9 such IDs reported here (see Table 3).  Three of these
(NV369$=$73a, V216$=$74d, and NV379$=$91a) have membership
probabilities in the range 94--99 per cent and are thus very likely members.
A fourth, NV371$=$11b, has a membership probability of
100 per cent according to \citet{vanLeeuwen2000}.  Of the other four,
according to \citet{Bellini2009}, two (NV377$=$82b and NV410$=$63h)
are clear non-members (membership probability $=$ 0 per cent), and two others
are probably unassociated with the cluster
(membership probabilities are 15 per cent and 23 per cent for V210$=$73d and
V167$=$84d, respectively).  Optical counterparts that are unlikely
to be associated with \wcen\ are shown in parentheses in column 13 of Table 3.

\section{A SYMBIOTIC STAR IN OMEGA CENTAURI}

The second-brightest source in the ACIS-I field of view (94a) lies about
8.8~arcmin southwest of the cluster centre (see Figs. 1 and 2) with a flux
of \fx (0.5--4.5 keV) $=$ 1.0 \x\ $10^{-13}$ \ergscmsq. This
source was also detected in our first epoch of \textit{Chandra} data
\citep{Haggard2009}, with \textit{XMM-Newton} \citep[][their source $\#$1]{Gendre2003}
and with \textit{ROSAT/PSPC} \citep[][their source $\#$11]{Johnston1994}, but
has not previously been identified optically.  In the \textit{XMM-Newton} study it was
the brightest and hardest source, and was variable on time-scales of
minutes to hours \citep[][see their Fig.~6]{Gendre2003}.

The position of this source coincides closely with a Population II
carbon star that was identified by \citet[][star 0055]{Harding1962} -- the 
first CH star to have been found in a globular cluster.  The
star, which lies near the tip of the RGB at V $=$ 11.49 and B--V $=$
1.74 \citep{Harding1962}, was later catalogued by
\citet{vanLeeuwen2000} as Cl\* NGC~5139 LEID~52030 at R.A. $=$
13\hours26\minutes01\secondspt61, Dec. $=$ $-$47\degrees33\amin05\secspt7.
This is 0.34 arcsec from the
\textit{Chandra} position for source 94a, inside the 95 per cent confidence radius of
0.55 arcsec (see Table 3).  The star was shown to be a radial-velocity
member of the cluster by \citet{Mayor1997} and a proper-motion member
at 99 per cent confidence by \citet{vanLeeuwen2000}.  It has since been
included in studies of mass-loss and dust production among giants in
\wcen\ \citep{vanLoon2007, McDonald2009}.

We propose that this object is a symbiotic star in \wcen.  CH stars in
the field are known to have compact binary companions that are
typically white dwarfs \citet{McClure1984, McClure1990}.  If the WD is
accreting from the giant's wind then it will appear as a symbiotic
star \citep{Kenyon1986}.  We show in Section 7 that its X-ray properties are
similar to several symbiotic stars recently observed with Suzaku
\citep{Nunez2016}.  Details concerning this star will appear
in a forthcoming paper.

\section{X-RAY CMD AND BRIGHT-SOURCE ANALYSIS}

Further clues to the nature of the X-ray sources in \wcen\ can be
gleaned from an examination of the cluster's X-ray colour--magnitude
diagram (CMD) shown in Fig.~7a.  Here we plot the wide-band (0.5--6.0
keV) flux vs.\ X-ray colour (Xcolour $=$ 2.5\x\ log[X$_{soft}$/X$_{hard}$]); the
corresponding \lx\ is indicated on the left-hand side assuming a
distance of 5.2 kpc.  Objects with known or suggested optical
counterparts are indicated with coloured symbols.  Solid and open
symbols signify cluster members and non-members, respectively.  For
comparison, we also plot hardness ratios for several spectral models,
using the known cluster absorption unless otherwise indicated.
Thermal plasma models (\texttt{vmekal} in XSPEC) appear at the top and
power-law models appear at the bottom; the vertical position is
arbitrary.  The effect of increasing the hydrogen column is indicated
for several values of \nH\ for a 10 keV thermal plasma spectrum.
Finally, hydrogen atmosphere neutron star models are plotted for a
range of temperatures, assuming a 10 km radius.  In Fig.~7b we show the
same diagram, but with symbols indicating location relative to the
cluster centre.  Errors on a subset of the points are also shown to
illustrate how the uncertainties depend on location within the plot.

\begin{figure}
    \centering
    \begin{subfigure}[b]{0.5\textwidth}
        \includegraphics[width=0.95\textwidth]{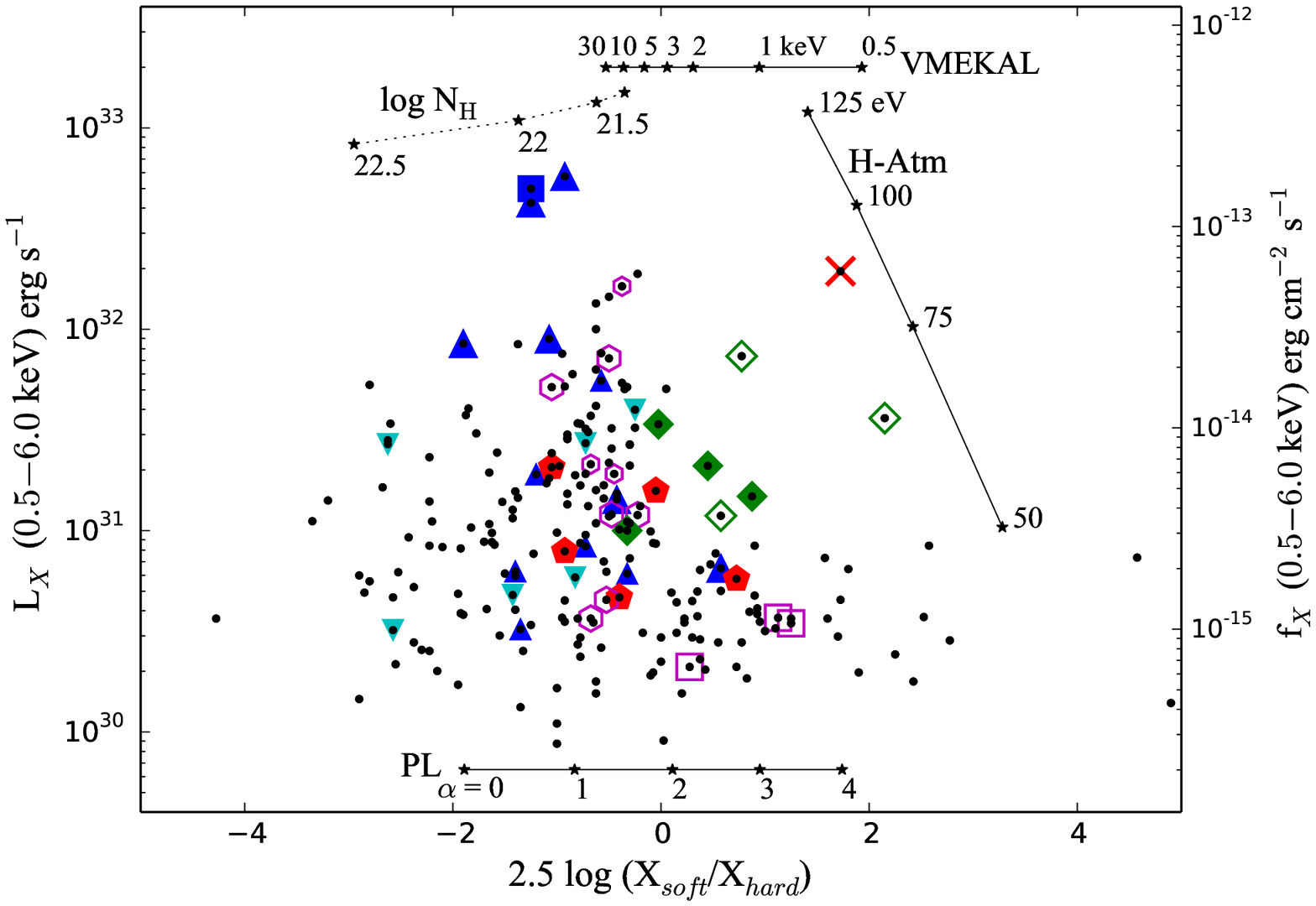}
        \caption{}
    \end{subfigure}
    
    \begin{subfigure}[b]{0.5\textwidth}
        \includegraphics[width=0.95\textwidth]{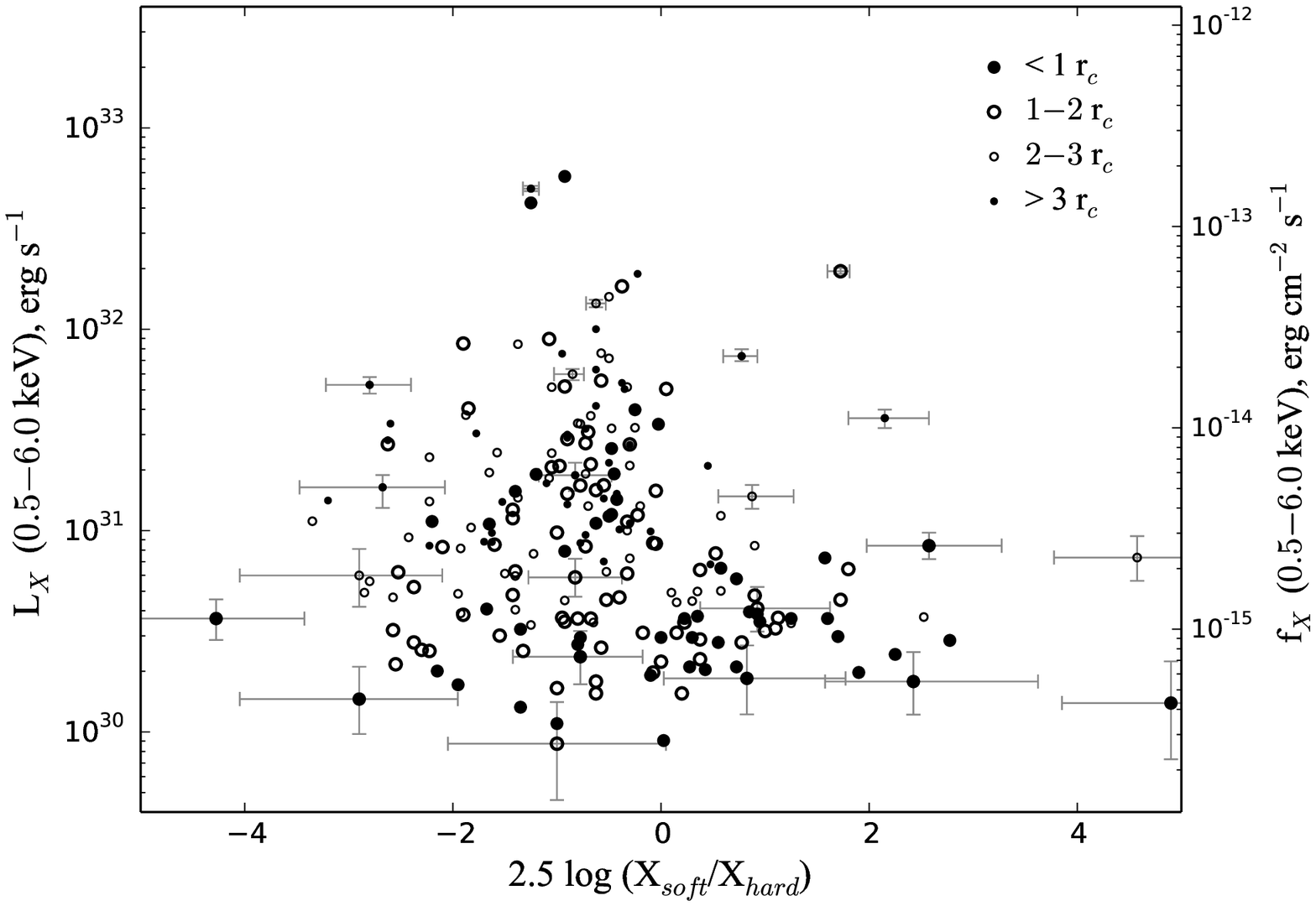}
        \caption{}
    \end{subfigure}
    \caption{(a) An X-ray `colour--magnitude' diagram for 221 \textit{Chandra}
          sources that have non-zero fluxes in both the hard ($1.5-6.0$~keV)
          and soft ($0.5-1.5$~keV) bands.  Symbols as in Fig.\ 6. Hardness
          ratios for several spectral models are shown using the known cluster
          absorption.  Thermal plasma models (\texttt{vmekal} in XSPEC) are shown near the top
          of the figure; power-law models appear close to the bottom (vertical position is arbitrary).
          The effect of increasing the hydrogen column is indicated for several
          values of \nH\ for a 10 keV thermal plasma spectrum. Hydrogen atmosphere
          neutron star models are plotted for a range of temperatures assuming a
          10 km radius.\\          
          (b) Same as top panel, but with symbols indicating location of sources
          in the cluster. Confidence intervals (68 per cent) on the X-ray colour values
          and wide-band fluxes calculated using the Bayesian estimation method outlined
          in \citet{Park2006} are shown for a representative sample of sources.}
\end{figure}

The three brightest X-ray sources -- two CVs and the CH star -- all have
hard X-ray colours suggestive of internal absorption.  The fainter CVs
and CV candidates display a significantly larger spread in colours and
are softer on average than the brightest CVs.  The known AGN have
X-ray colours similar to the CVs, highlighting the importance of
optical identifications.  The X-ray sources associated with variable
stars from \citet{Kaluzny1996, Kaluzny2004} have relatively soft X-ray
colours, similar to those of the three known foreground stars,
consistent with their probable coronal nature.  X-ray colours of the
RGB/SGB-a stars are intermediate between the accretion sources and the
coronal sources.

For further insight into the brightest cluster X-ray sources, we
performed spectral fitting for 14 sources whose optical counterparts
suggest they are cluster members.  We selected sources with at least
70 counts in the 0.5--6.0 keV band.  This includes the CH star, nine
CVs and CV candidates, two stars that lie on the anomalous RGB/SGB,
and two of the Kaluzny variables.  Since the spectrum of the qLMXB has
already been reported elsewhere \citep[][see also \citet{Rutledge2002}]{Heinke2014}, we do not discuss
it here.  

We began by combining the spectra from the two Cycle 13 ObsIDs using
the HEAsoft tool \textsc{addspec}.  We then fitted the combined
spectra using HEAsoft/Xspec and C-statistics \citep{Cash1979}.  We fit
\texttt{vmekal} models (\texttt{tbabs$*$vmekal}), specifying abundances appropriate for
the dominant [Fe/H] $= -1.5$ stellar population in \wcen.  
For the faintest sources we fixed the column at the cluster value of
\nH\ $=$ 9$\times$10$^{20}~$cm$^{-2}$, while for those with at least
100 counts we allowed the \nH\ value to vary, which yields a more
meaningful constraint on kT.
Best-fitting parameters, including 90 per cent confidence intervals, are reported in Table 2.

\input table3_FINAL.tex

For the CH star and five of the nine brightest CVs and CV candidates
no satisfactory fit could be obtained at the cluster \nH\ value,
corroborating the X-ray CMD results.  For these sources (94a, 13c,
13a, 54h, 41d, and 54b), the cluster \nH\ value is excluded at 90 per cent
confidence or more.  The required \nH\ values for these sources lie
in the range 4\x 10$^{21}$--2\x 10$^{22}$~cm$^{-2}$, similar to what
\citet{Heinke2005} found for the most strongly absorbed sources in
47~Tuc.  Internal absorption was also required to obtain a good fit
for source 34b, an RGB/SGB-a star.  We discuss the significance of
these findings below (see Section 8).  
For four other sources for which \nH\ was allowed to vary (43h, 22c, 31a, and
11b), no internal absorption was required to obtain a satisfactory fit.
All the CVs and CV candidates were well fit with plasma
temperatures in the range 6$-$30 keV which is typical of CVs
\citep{Mukai2017}.  The CH star, with a plasma temperature of 7 keV,
appears softer than the two CVs with comparable luminosities.  Its
temperature, luminosity, and enhanced \nH\ values are all within the
range of properties deduced for a set of symbiotic stars observed with
Suzaku \citep{Nunez2016}.  The two RGB/SGB-a stars and one of the
Kaluzny variables have lower plasma temperatures in the range 2$-$3
keV which is more typical of coronal sources \citep{Dempsey1993,
  Dempsey1997}.  Sample spectra are shown in Fig.~8.
  
\begin{figure}
    \centering
    \begin{subfigure}[b]{0.23\textwidth}
        \includegraphics[width=\textwidth]{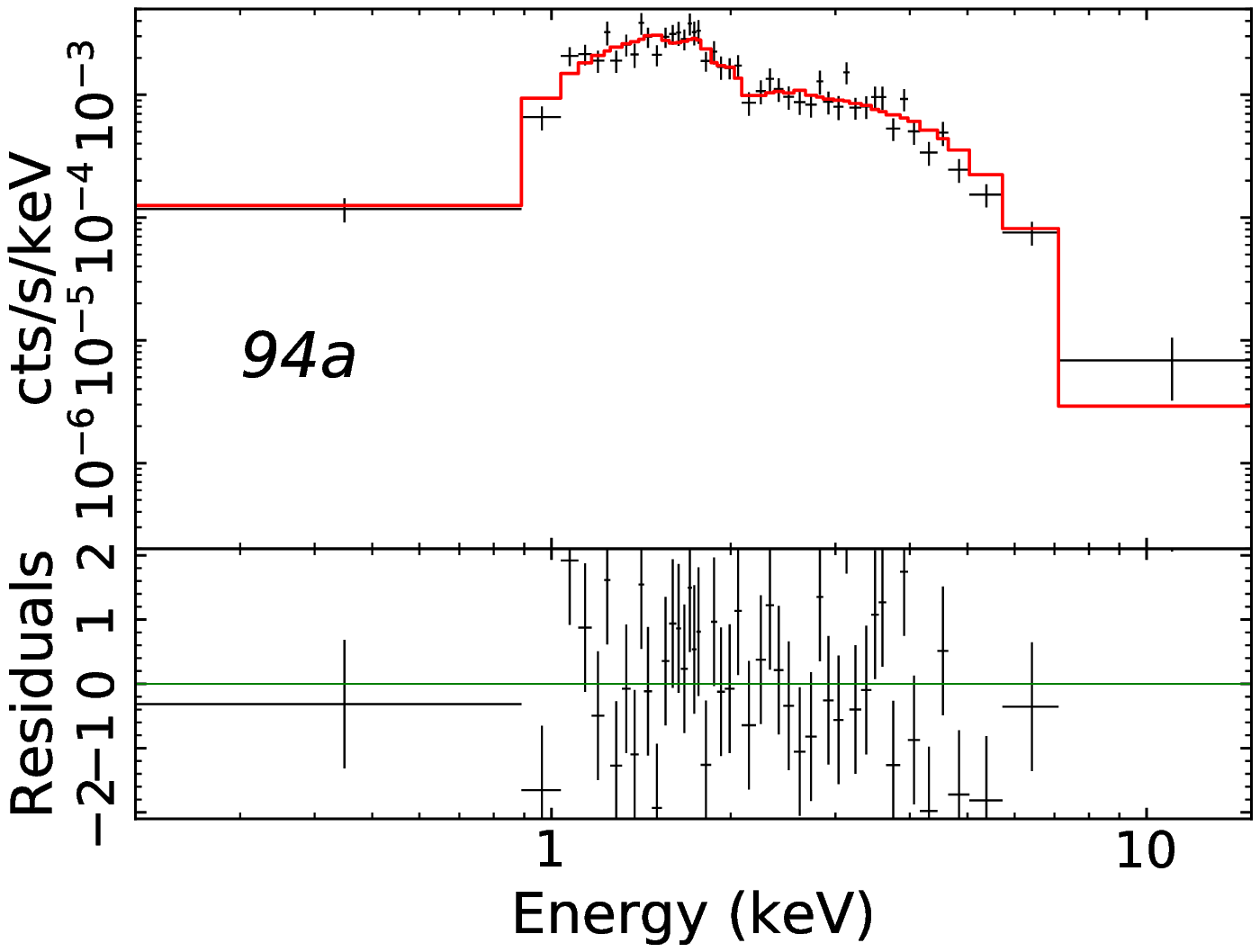}
        \caption{}
    \end{subfigure}
    ~
    \begin{subfigure}[b]{0.23\textwidth}
        \includegraphics[width=\textwidth]{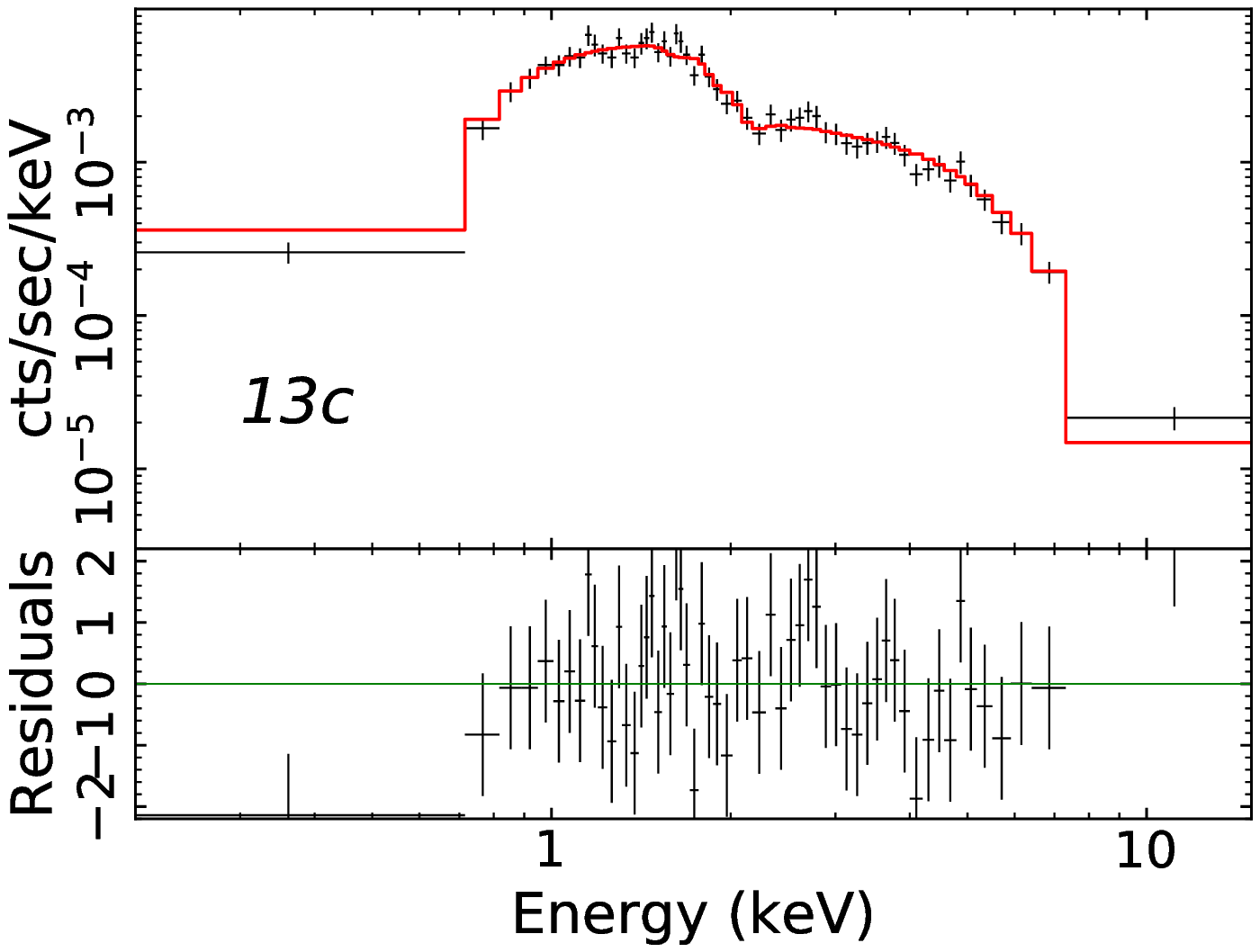}
        \caption{}
    \end{subfigure}

    \begin{subfigure}[b]{0.23\textwidth}
        \includegraphics[width=\textwidth]{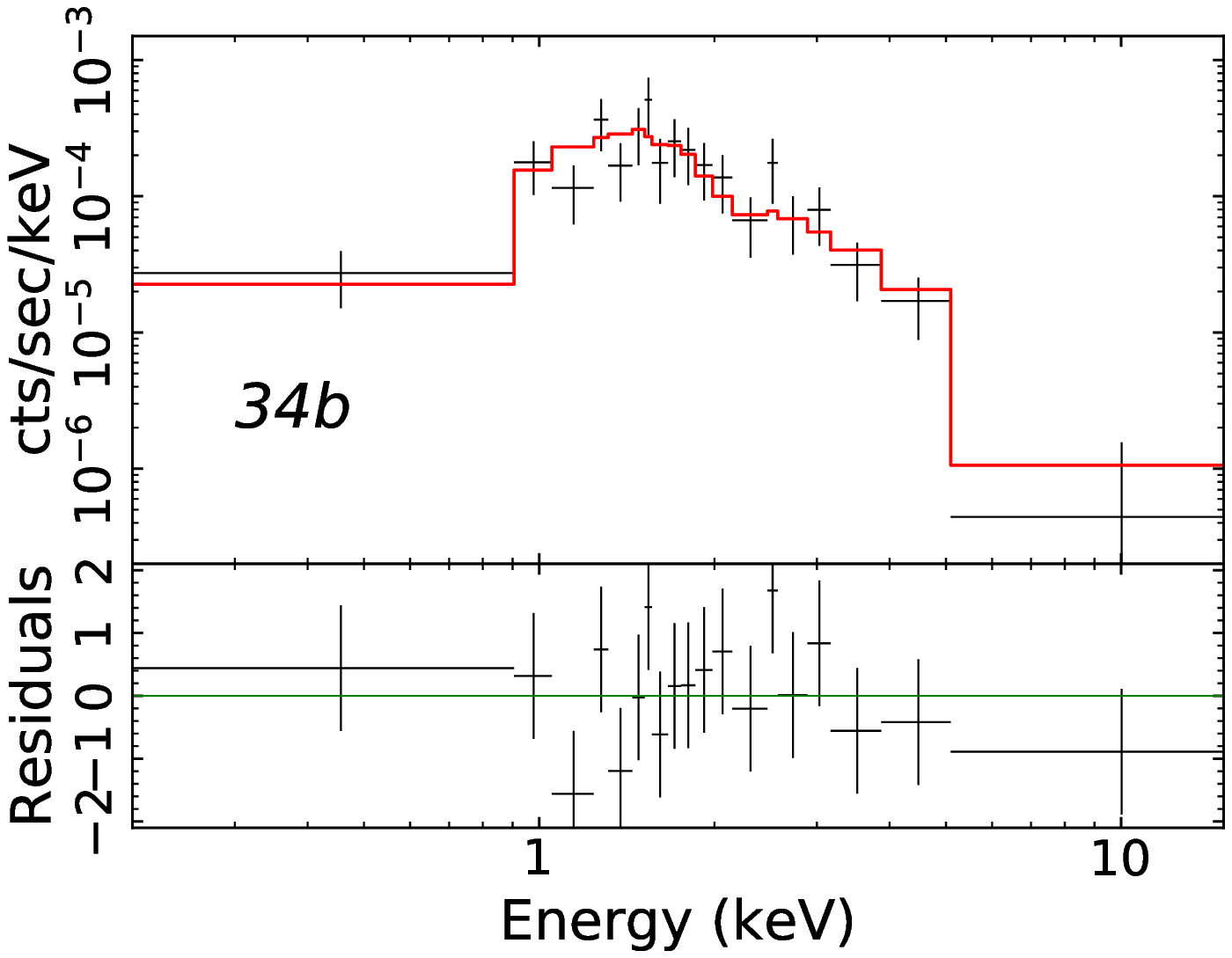}
        \caption{}
    \end{subfigure}
    ~
    \begin{subfigure}[b]{0.23\textwidth}
        \includegraphics[width=\textwidth]{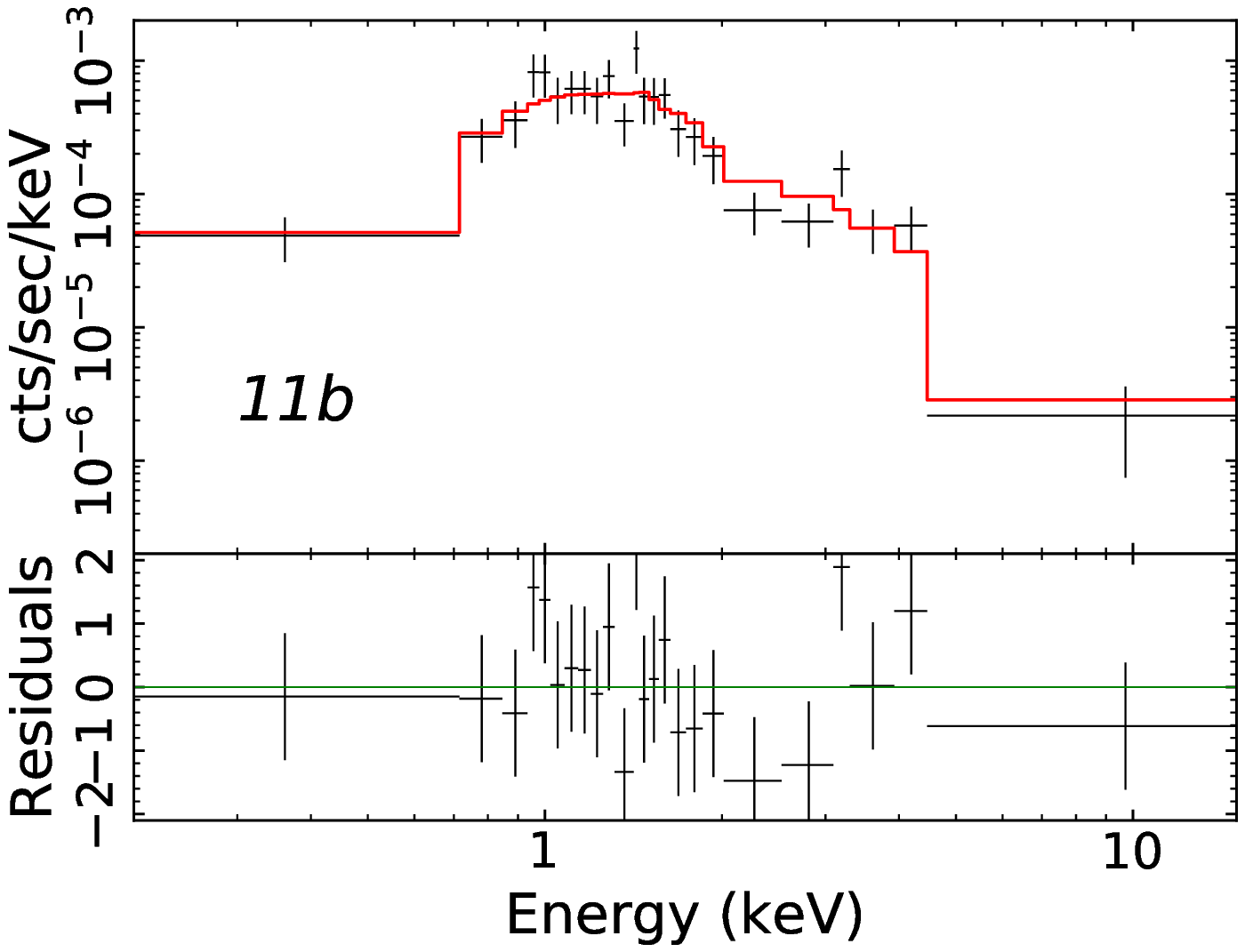}
        \caption{}
    \end{subfigure}
    \caption{Spectra of four X-ray sources in \wcen: (a) the CH star (94a), (b) the brightest CV (13c), (c) the brightest RGB/SGB-a star (34b), and (d) the brightest of the Kaluzny variables (11b).  Solid red lines in upper panels represent fits of \texttt{vmekal} models to the Cycle 13 spectra made using a maximum-likelihood procedure (see text).  Spectra are binned for plotting purposes only.  Lower panels show the residuals: (data$-$model)/error.}
\end{figure}

We also searched the X-ray lightcurves of these 14 bright sources for
signs of variability.  We used CIAO's \textsc{glvary} tool to search
for variability within each Cycle 13 ObsID.  
The CIAO \textsc{glvary} tool applies the Gregory-Loredo variability algorithm
\citep{Gregory1992} which looks for significant changes between events
in different time bins, and assigns a variability index 
and a probability that the flux from the source region is time-variable.
We also searched for variations between the two ObsIDs by simultaneously fitting spectra
extracted from each while allowing the normalization to differ.  The
CH star shows clear variability in the longer of the two ObsIDs
(\textsc{glvary} index $=$ 6), with evident flaring activity (see
Fig.~9).  Several other sources show hints of variability up to
\about 50 per cent, but none with greater than 90 per cent confidence.  Additional
spectral and variability analyses are underway and will be presented
elsewhere.

\begin{figure}
\centering
  \includegraphics[width=\linewidth]{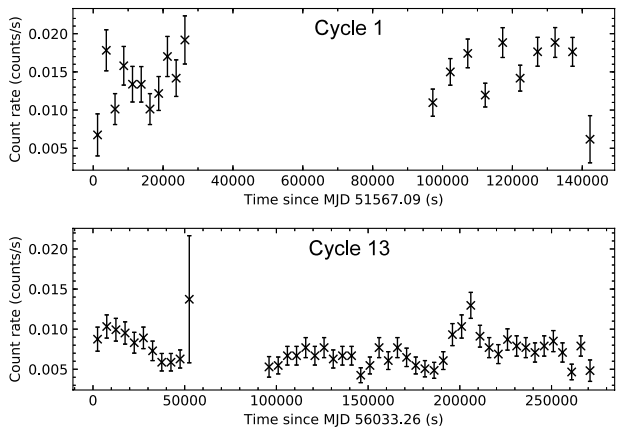}
  \caption{X-ray lightcurve for source 94a, a proposed symbiotic star in \wcen.  Cycle 1 and Cycle 13 data are shown in top and bottom plots, respectively,
each of which are divided into two separate observation periods (ObsIDs).
Significant flaring activity can be seen in the second Cycle 13 ObsID.}
\end{figure}

Among the sources in the X-ray CMD that have yet to be identified
optically, many have rather hard X-ray colours (e.g.\ requiring more
than 3\x 10$^{21}$~cm$^{-2}$ of additional absorption).  This is
unlikely to be the result of differential interstellar absorption,
which should be relatively small across our field \citep[of order
  10 per cent;][]{Bellini2017}.  However it is not unexpected given that the
majority of the unidentified sources are likely to be AGN (see Section
4), which have hard spectra and frequently show internal absorption
\citep{Hasinger2001, Kim2004}.  Of the 45 sources with optical IDs
thus far, 29 appear to be members, leaving \about 30 member sources
yet to be identified out of more than 180 with no IDs at present.

\section{DISCUSSION}

These \textit{Chandra} observations bring the total number of X-ray
sources known in and toward Omega Centauri to 275, a factor of \about
1.5 increase over those previously known \citep{Haggard2009}.  Of the
233 sources in the present data, we estimate that \about $60\pm 20$ are
associated with the cluster.  The remainder are primarily AGN, with a
smaller contribution from foreground stars.  In projection,
approximately 30 of the cluster sources reside in the core; a similar
number lie in a 1--2\rc\ annulus surrounding the core.  The sources
range in luminosity from \lx\ \about 1\x $10^{30}$ to 6\x $10^{32}$~\ergs\
in the 0.5--6.0~keV band, or \about 3\x $10^{29}$ to 2\x
$10^{32}$~ \ergs\ in the 0.5--2.0~keV band.  Relatively few member
sources lie outside 2\rc, but our knowledge of this population is more
limited owing to the drop in sensitivity off-axis.  In the 0.5--2.0~keV
band, these observations should be complete to
\lx\ $\simeq\ 10^{30}$~\ergs\ in the cluster core and to
\lx\ $\simeq\ 10^{31}$~\ergs\ over the full ACIS-I field of view
(see Fig.~6).\footnote{These \lx\ values increase by a factor of \about 2--3 if
we consider the 0.5--6.0~keV band instead.}

Optical identifications reported in previous studies, combined with
four new IDs presented here, bring the number of X-ray sources in
\wcen\ with secure or promising optical counterparts to 45.  These
include 29 cluster members, 20 of which are accretion-powered: one
qLMXB, 18 CVs and CV candidates, and one newly identified symbiotic star.
Nine others are most likely coronal sources: five stars that lie along
\wcen's anomalous giant and subgiant branch and four variable stars.

Cataclysmic variables dominate the X-ray source population in
\wcen\ for which optical IDs exist.  
Extrapolating their fraction (18/29) to the total
X-ray source population suggests a total of \about $40\pm 10$ CVs in
\wcen\ with \lx (0.5--2.0~keV) $\ge$ 10$^{30}$~\ergs, compatible with
the upper end of an earlier estimate made by \citet{Haggard2009}.
Given the greater ease with which accretion vs. coronal sources can
typically be identified in optical searches, this may be an
overestimate.  Still, it seems probable that at least \about 30 CVs
are present in \wcen\ given the number already known, the difficulty
of obtaining optical identifications in crowded fields like
\wcen, and the fact that optical follow-up has yet to be undertaken
for the 95 newly-identified X-ray sources.

Theoretical work shows that CVs in globular clusters should form
through two distinct channels.  Primordial binaries that would have
given rise to CVs in the field can, under
favourable conditions, similarly produce CVs in a GC
\citep{Davies1997, Ivanova2006}.  Dynamical interactions taking place
primarily in cluster cores provide a second channel.  The dominant
mechanism for CV formation in this case involves an exchange
interaction in which a typically heavy white dwarf is exchanged into a
primordial binary consisting of two main-sequence stars.  This channel
is favoured in clusters with high-density and/or large cores which
generate high rates of stellar interactions \citep{Ivanova2006,
  Belloni2016, Hong2017}.

Omega Cen presents an interesting case in the context of these two CV
formation channels.  Given the sheer number of stars it contains, it is
likely to have formed with many of the primordial binaries that could
give rise to CVs, and its modest central density \citep[$\rho _0$
  \about\ 3\x $10^3$~\msun/pc$^3$;][]{Pryor1993} favours the survival
of such systems.  At the same time, because it has a very large core
(r$_c$ $\simeq$ 3.9~pc), the overall rate of stellar interactions is
high enough that CVs are also expected to form via dynamical
interactions \citep{Verbunt1988, Davies1995}.  In addition, \wcen's
half-mass relaxation time is sufficiently long \citep[1.2
  \x\ 10$^{10}$~yr;][]{Harris2010} that, in contrast to most GCs with
significant rates of stellar interactions, \wcen\ is far from being
relaxed, as demonstrated by its lack of significant mass segregation
\citep{Anderson2002}.  This further favours the survival of primordial
binaries that could give rise to CVs, since they are less likely to
have sunk to the more perilous central regions of the cluster despite
their larger-than-average masses.  Moreover, in 
contrast to CVs in most other GCs whose radial distribution reflects
their masses \citep[e.g.][]{Cohn2010}, the distribution of CVs in
\wcen\ should be more indicative of where they formed, regardless of
origin.

The radial distribution of optically identified CVs in \wcen\ can be
seen in Fig.~10, in which we plot X-ray luminosity as a function of
radius in the cluster.\footnote{\textit{HST}-based searches for optical
  counterparts to date have focused on the region interior to
  \rh\ \citep{Cool2013}; less is known about CVs outside
  the half-mass radius.} More than half of the known CVs lie outside
0.5\rh; there is no sign that CVs preferentially reside in the denser
central region of the cluster.  This is in contrast to other clusters
in which significant numbers of CVs have been observed.  
In NGC~6397, NGC~6752, and 47~Tuc, CVs are clearly concentrated toward
the cluster centres and dynamical interactions appear to be implicated in the formation
of at least some of these systems \citep{Cohn2010, Lugger2017, RiveraSandoval2018}.
The markedly different radial distribution observed here suggests that
the bulk of the CVs in \wcen\ have their origins as primordial
binaries, which have survived by inhabiting the regions outside the
cluster core.

\begin{figure}
\centering
  \includegraphics[width=\linewidth]{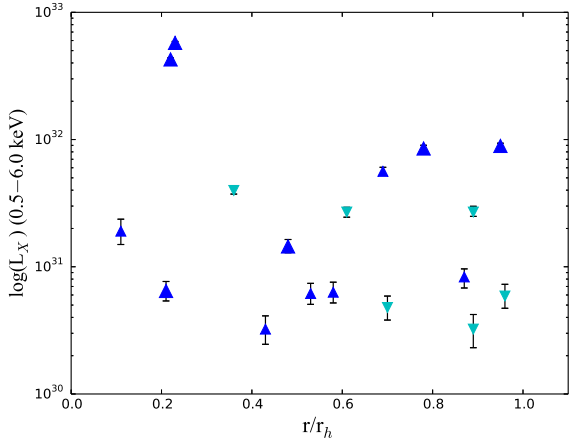}
  \caption{X-ray luminosity of CVs and CV candidates as a function of radius in \wcen,
in units of the half-mass radius (5.0~arcmin).  Symbols as in Fig.\ 6.}
\end{figure}

While it appears likely that \wcen's CV population is dominated by 
systems originating from primordial binaries, it is interesting to ask if
there are indications that any of the CVs in the cluster could be of
dynamical origin.  The rate of stellar interactions in \wcen\ is
expected to be comparable to the collapsed-core cluster NGC~6397
\citep{Bahramian2013} whose central density is \about 60 times higher
than \wcen's \citep{Pryor1993}, but whose core is far smaller.  Among
the 15 CVs in NGC~6397, several appear to be of dynamical origin:
their concentration toward the cluster centre implies high masses that
are likely the result of exchange collisions, and their high
luminosities are indicative of youth \citep{Cohn2010}.  Of these, the
four most luminous CVs, with \lx (0.5--6.0~keV) $>$ 5\x $10^{31}$~\ergs\
lie in or very near the cluster core.  In \wcen, the two most
luminous CVs lie well inside the core (\rc $=$ 2.6~arcmin), just
1.1~arcmin and 1.2~arcmin from the cluster centre, respectively.
While the numbers are small, this is at least a hint that the
dynamical formation channel is operating in \wcen.  The probability
that the two most recently formed primordial CVs would both lie so close to the
cluster centre by chance is \about 1 per cent judging from the number of
stars inside 1.2~arcmin vs. inside 2\rc.\footnote{These figures were
  determined based on turnoff stars in the ACS/WFC data described by
  \citet{Cool2013}.}  Moreover, the rate of interactions involving
heavy remnants may be highest in the inner region of the core where
these sources lie.  Given \wcen's \about 4~Gyr central relaxation
time, such remnants, which are more massive than average stars in the
core and formed early in the cluster's history, would have had time to
form a relaxed and thus more centrally concentrated distribution than
the main-sequence stars in the core.

While \wcen\ harbours a substantial population of CVs, their numbers
appear low by comparison to the field.  Based on a sample of 20
non-magnetic CVs with \lx (0.5--2.5~keV) \about 10$^{30}$--10$^{32}$~\ergs,
\citet{Pretorius2012} deduce a CV space density of
4$^{+6}_{-2}$ \x 10$^{-6}$~pc$^{-3}$ (1 $\sigma$ confidence interval) in the solar neighborhood.  The
range of X-ray luminosities we sample in \wcen\ is very
similar.
The estimated population of \about $40\pm10$
CVs with \lx (0.5--2.0~keV) \simgreat\ 10$^{30}$~\ergs\ translates
to a space density of \about 1.3 \x 10$^{-6}$~pc$^{-3}$ considering
the cluster mass (\about $3\times 10^6$~\msun) and assuming 0.1~\msun/pc$^{-3}$
in the solar neighborhood.  
This is a factor of three
below the space density in the field, although it is within 2 $\sigma$ of the field value.
However, the field estimate
takes account of non-magnetic CVs only,\footnote{Including magnetic
  CVs in this estimate increases the numbers by a factor of \about 1.2
  \citep{Pretorius2013}.} while there are hints that several of the CVs
in \wcen\ may be magnetic (see below), which would increase the
apparent discrepancy.  
We conclude that \wcen\ has fewer CVs per unit mass than the field.\footnote{This conclusion
applies to CVs with \lx\ $>$ 10$^{30}$~\ergs.  Fainter CVs are likely to be present in the field 
\citep{Pretorius2012} and presumably also in \wcen, but at present are not well constrained.}
It seems unlikely that the central densities in \wcen\ are
sufficiently high to destroy CVs, or even (wider-orbit) CV
progenitors; for instance, \citet{Ivanova2006} find a similar or
larger number of CVs per unit mass in their simulation of a cluster
of similar density to \wcen, compared to the field.  The reduction in
CVs per unit mass might be due to a lower initial binary fraction in
\wcen\ than the field, and/or to effects of lower metallicity and
larger age.  These issues will be studied in more detail in Heinke et
al.\ 2018, in prep.

Additional insight into the CVs in \wcen\ comes from spectral fits to
the brightest sources in the cluster (see Table 2).  Hydrogen columns
significantly above the cluster value were required to obtain good
fits for six of the nine CVs in this list.  The \nH\ values implied
for these sources range from \about 4\x 10$^{21}$~cm$^{-2}$ to 2\x
10$^{22}$~cm$^{-2}$.  Comparably large \nH\ values were found by
\citet{Heinke2005} for four CVs in 47~Tuc (see their Fig.~17), three
of which are known to be eclipsing systems.  Thus it is possible that
at least some of the high-\nH\ CVs in \wcen\ are also high-inclination
systems.  From a statistical point of view, however, it would be
surprising if 2/3 of the brightest CVs in \wcen\ were all edge-on.
An alternative explanation could be that at least some are
magnetic.  Intermediate polars, also known as DQ~Her systems, are
often found to require internal absorption \citep{Norton1989,
  Patterson1994, Mukai2017}.  These moderately magnetic systems are
also brighter on average than non-magnetic systems
\citep{Pretorius2013}, which could help explain why so many of the
bright CVs in \wcen\ display this feature.

One of the sources associated with an RGB/SGB-a star also shows
enhanced absorption over the cluster value (see Table 2).  These are
stars that, in colour-magnitude diagrams, lie along the metal-rich
anomalous subgiant and giant branches in \wcen\ \citep{Cool2013},
hence the designation.  It is currently unknown, however, whether they
have chemical compositions that actually make them part of this
population.  If instead their metallicities are characteristic of one
of the more metal-poor populations in \wcen, then they are
sub-subgiants \citep{Mathieu2003, Geller2017a, 
  Geller2017b, Leiner2017} or, in the case of brighter members of the group, red
stragglers.  In this context the enhanced emission associated with the
possible sub-subgiant (SSG) 34b is interesting given that enhanced absorption has
been found to be associated with a number of other such systems
\citep{Mathieu2003} and multiple mechanisms put forward to
explain SSGs involve mass loss and/or mass transfer
\citep{Leiner2017}.  Spectroscopic follow-up on this and other members
of this population in \wcen\ is needed to distinguish between these
possibilities.

A class of objects that appears conspicuously absent from \wcen\ is
millisecond pulsars (MSPs).  In contrast to 47~Tuc, which hosts 25
known MSPs \citep{Freire2017}, none has yet been detected in \wcen.
This may not be surprising, considering the role that stellar
interactions are likely to play in the formation of MSPs in globular
clusters \citep{Heinke2010}, and the fact that the rate of interactions in
\wcen\ is more than an order of magnitude below that of 47~Tuc \citep{Bahramian2013}.
NGC~6397, with an interaction rate comparable to that of \wcen, has just
one known MSP.  On the other hand, gamma-ray emission from the cluster
centre detected with \textit{Fermi/LAT} \citep{Abdo2010} hints at the presence
of a small population of MSPs, formally estimated at $19\pm 9$.  Thus
it is of interest to ask whether the present X-ray observations can
tell us anything about the possibility that some MSPs could exist in
\wcen\ and have somehow eluded detection.  The known MSPs in 47~Tuc occupy
a relatively small region of the X-ray CMD
\citep[see Fig.~10 of][]{Heinke2005}; the large majority have \lx\ $=$
10$^{30}$--10$^{31}$~\ergs\ and X-ray colours that place them between
the 0.5 and 3.0 keV \texttt{vmekal} models.\footnote{These colour boundaries are
  chosen for convenience of comparing to X-ray sources in \wcen, not
  because MSPs are well fit by \texttt{vmekal} models.}  The equivalent region
in \wcen\ (see Fig.~7a), contains \about 40 sources, all but 5 of which
are unidentified.  Thus the present observations leave open the
possibility than some MSPs could exist in \wcen.  However, given that
multiple source types including CVs and active binaries also occupy
this region of the X-ray CMD, radio detections are required to
determine whether any of these sources are in fact MSPs.

Finally, we identify the first candidate symbiotic star in a globular
cluster with 94a, the second brightest X-ray source in \wcen.  In
section 6 above, the nature (a carbon-rich red giant, or CH star) and
membership in the cluster of the optical counterpart are secured.  In
section 7, the X-ray spectrum of 94a is reasonably fit with a plasma
temperature of 5 keV and intrinsic absorption of $7\times10^{21}$~cm$^{-2}$.
Although the first well-studied symbiotic stars tended to
show quite soft spectra (e.g. \citet{Muerset1997} using \textit{ROSAT} data),
recent studies of symbiotic stars have revealed that many have hard
X-ray spectra (temperatures 5-50 keV), with substantial
(\about $10^{22}$--$10^{23}$~cm$^{-2}$) absorption \citep[e.g.][]{Luna2013}.
Thus, 94a's X-ray spectrum is quite consistent with those of symbiotic
stars.  No high-temperature emission lines have yet been identified
from this star.  However, as laid out clearly by \citet{Mukai2016} and also
\citet{Hynes2014}, \citet{vandenberg2006}, and \citet{Munari2002},
symbiotic stars will only show strong optical
emission lines if a very large ($>4\times10^{34}$~\ergs) ionizing
source is present.  Thus, most symbiotic stars, including many with
substantial ($>10^{32}$~\ergs) X-ray luminosity, will not show strong
optical emission lines.  So from the optical perspective as well,
94a's properties are in line with our current understanding of
symbiotic stars.  We note that the relative rarity of symbiotic stars
in globular clusters, compared to closer binaries, is a natural
consequence of the wider orbits of symbiotic stars.  Such binaries
would be disrupted in the dense cores of globular clusters, so
symbiotic stars should be produced only from primordial binaries in
the low-density haloes of (preferentially massive) globular clusters.

\section{SUMMARY}

Analysis of a deep \textit{Chandra} exposure of the globular cluster Omega
Centauri has revealed 233 X-ray sources in the ACIS-I field of view,
of which 95 are newly reported here.  An estimated $60\pm 20$ of these
sources are cluster members, the remaining being primarily AGN.  Among
45 sources with firm or tentative optical identifications are 18 CVs
and CV candidates.  Extrapolating from these, we estimate that the
cluster contains \about 30--40 CVs with \lx\ \simgreat\ 10$^{30}$~\ergs,
a factor of about three fewer than would be expected if
\wcen\ produced CVs at the same rate as the field.  In contrast to
other GCs with significant populations of known CVs, the majority of
CVs in \wcen\ lie outside the cluster core.  Given \wcen's very long
half-mass relaxation time, this strongly suggests that a majority of
\wcen's CVs have evolved from primordial binaries.  The two brightest
CVs lie close to the cluster centre, hinting that a dynamical
formation channel may also be active.  Spectral analysis shows that
five of the nine brightest CVs have significant internal absorption
indicative of a possible magnetic nature.  An X-ray CMD contains
numerous unidentified sources with colours and luminosities typical of
MSPs.  Radio identifications are needed to determine whether any of
these objects are MSPs and could help explain the gamma-ray emission
seen from the central regions of the cluster \citep{Abdo2010}.
Finally, we identify the second-brightest X-ray source present as a
symbiotic star, the first such binary system found in a globular
cluster.

\section{ACKNOWLEDGEMENTS}

We thank B.~Luo for sharing results of the 2 Msec \textit{Chandra}
Deep Field South study.  Support for this work was provided by the
National Aeronautics and Space Administration through \textit{Chandra}
Award Number GO0-1040A issued by the \textit{Chandra X-ray Observatory
  Center}, which is operated by the Smithsonian Astrophysical
Observatory for and on behalf of the National Aeronautics Space
Administration under contract NAS8-03060.  DH acknowledges support
from a Natural Sciences and Engineering Research Council of Canada
(NSERC) Discovery Grant and a Fonds de recherche du Qu\' ebec - Nature et
Technologies (FRQNT) Nouveaux Chercheurs Grant.  CH is supported by an
NSERC Discovery Grant and an Accelerator Supplement.



\bibliographystyle{mnras}
\bibliography{man.bib}

\clearpage

\input table1_FINAL.tex


\label{lastpage}
\end{document}

%% file: table2_FINAL.tex
\begin{table}
\centering
  \caption{Adopted aperture radii and background annuli.}
  \vskip 0.1in
  \begin{tabular}{ccc}
        \hline\hline
        Offset & Aperture Radius & Background Annulus\\
        (arcminutes)   & (pixels)        & (pixels)             \\
        \hline
        $<$4           & 4               & 15$-$30              \\
        4$-$5          & 6               & 15$-$30              \\
        5$-$6          & 8               & 15$-$30              \\
        6$-$7          & 10              & 30$-$45              \\
        7$-$9.5        & 12              & 30$-$45              \\
        $>$9.5         & 14              & 30$-$45              \\
  \end{tabular}
  \label{table:aperture-sizes}
\end{table}

%% file: table3_FINAL.tex
\begin{table*}
  \caption{Spectral Fits of Brightest Sources in Omega Cen.}
\vskip 0.1in
  \bgroup
  \def\arraystretch{1.4}
   \begin{tabular}{cccccccc}
        \hline\hline
        Source & Type      & Counts         & \nH$^\dagger$         & kT & L$_{X,abs}$$^\ddagger$ & L$_{X,unabs}$$^\ddagger$ \\
               &           & (0.5--6.0 keV) & (10$^{20}$ cm$^{-2}$) & (keV) & (10$^{30}$ erg s$^{-1}$) & (10$^{30}$ erg s$^{-1}$) \\
        \hline
	94a    & CH star   & 1092           &  72$^{+16}_{-15}$      & 6.8$^{+3.3}_{-1.7}$      & 466$^{+25}_{-26}$ & 662$^{+52}_{-52}$ \\
	\\
	13c    & CV        & 2252           &  16$^{+6}_{-6}$      & 34.0$^{+45.4}_{-13.9}$   & 570$^{+21}_{-21}$ & 627$^{+27}_{-27}$ \\
	13a    & CV        & 2053           &  51$^{+8}_{-8}$      & 14.4$^{+7.9}_{-4.0}$     & 423$^{+16}_{-16}$ & 539$^{+27}_{-27}$ \\
	54h    & CV        &  440           &  37$^{+20}_{-18}$     & 19.2$^{+\infty}_{-10.5}$ & 98$^{+8}_{-8}$    & 119$^{+9}_{-10}$  \\
	41d    & CV        &  407           & 112$^{+35}_{-30}$    & 14.6$^{+\infty}_{-7.7}$  & 117$^{+9}_{-10}$  & 172$^{+14}_{-14}$ \\
	43h    & CV?       &  224           &  12$^{+16}_{-3^*}$    & 6.8$^{+8.2}_{-3.1}$      &  49$^{+5}_{-5}$   &  54$^{+6}_{-6}$   \\
	22c    & fbCV?     &  193           &  11$^{+15}_{-2^*}$   & 5.7$^{+5.6}_{-2.1}$      &  35$^{+4}_{-4}$   &  39$^{+4}_{-5}$   \\
	31a    & fbCV?     &  130           &  21$^{+25}_{-12^*}$   & 22.3$^{+57.6}_{-15.7}$   &  27$^{+4}_{-4}$   &  31$^{+4}_{-4}$   \\
	54b    & fbCV?     &  125           & 196$^{+84}_{-72}$    & 13.0$^{+\infty}_{-8.3}$  &  34$^{+5}_{-5}$   &  58$^{+8}_{-9}$   \\
	24c    & CV        &   75           &   (9)              & 5.7$^{+13.2}_{-2.7}$     &  12$^{+2}_{-2}$   &  13$^{+2}_{-3}$   \\
	\\
	34b    & RGB/SGB-a &   93           &  55$^{+41}_{-34}$    & 2.7$^{+4.4}_{-1.0}$      &  17$^{+3}_{-3}$   &  25$^{+4}_{-4}$   \\
	32f    & RGB/SGB-a &   76           &   (9)              & 2.2$^{+1.6}_{-0.7}$      &  12$^{+2}_{-2}$   &  13$^{+2}_{-3}$   \\
	\\
	11b    & NV371     &  177           &   9$^{+7}_{-0^*}$    & 3.2$^{+1.2}_{-0.7}$      &  26$^{+3}_{-3}$   &  29$^{+3}_{-4}$   \\
	73a    & NV369     &   71           &   (9)              & 8.8$^{+47.6}_{-5.1}$     &  18$^{+3}_{-4}$   &  19$^{+4}_{-4}$   \\
   \end{tabular}
   \egroup
\tablecomments{$^\dagger$Parentheses indicate that the parameter was frozen during the fit; $^\ddagger$Luminosities are given in the $0.5-6.0$ keV band; $^*$Lower limits are below minimum limiting value for \nH\ (9$\times$10$^{20}~$cm$^{-2}$).}
\end{table*}

%% file: table1_FINAL.tex
\begin{deluxetable}{ccccccccccccc}
\rotate
\tabletypesize{\scriptsize}
\tablecolumns{13}
\tablewidth{0pc}
\tablecaption{Omega Cen X-ray Sources}
\tablehead{
           \colhead{}        & \colhead{}             & \colhead{}                & \colhead{} & \multicolumn{3}{c}{Counts Detected/Corrected}                                                              &   \colhead{}                                               & \multicolumn{3}{c}{$f_x$}                                                      &  \colhead{$f_x$ ratio$^e$} & \colhead{}  \\
                                                                                             \cline{5-7}
           \colhead{}        & \colhead{}             & \colhead{Error$^c$} & \colhead{Offset$^d$}       & \colhead{$X_{med}$} & \colhead{$X_{soft}$} & \colhead{$X_{hard}$} & \colhead{} &    \multicolumn{3}{c}{($10^{-16}$ erg cm$^{-2}$ s$^{-1}$)}    &  \colhead{2012/2000}       &  \colhead{Optical}  \\
                                                                                                                                                                                                                                                        \cline{9-11}
           \colhead{Src$^a$} & \colhead{Position$^b$} & \colhead{(arcsec)}         & \colhead{($r_c$)} & \colhead{0.5--4.5 keV}     & \colhead{0.5--1.5 keV}      & \colhead{1.5--6.0 keV}     & \colhead{$\log\left(\frac{X_{soft}}{X_{hard}}\right)$} & \colhead{0.5--4.5 keV} & \colhead{0.5--6.0 keV} & \colhead{0.5--2.0 keV}            & \colhead{0.5--6.0 keV}                  &  \colhead{ID$^f$}
          }
\startdata
11a			&		J132641.506$-$472832.70		&		0.49		&	0.39	&		11/10.0 		&		6/5.1   		&		6/5.7   		&		$-0.04$		&		4.3		&		5.9		&		3.9		&		1.37		&	 \ldots  \\
11b			&		J132641.021$-$472737.30		&		0.33		&	0.60	&		173/183.2		&		97/94.9 		&		80/96.5 		&		$-0.01$		&		79.6		&		104.3		&		51.4		&		1.47		&	 NV371  \\
{\bf 11d}		&		J132640.786$-$472832.80		&		0.46		&	0.43	&		14/13.6 		&		12/11.1 		&		2/1.4   		&		0.90		&		6.0		&		7.5		&		4.7		&		\ldots		&	 \ldots \\
{\bf 11e}		&		J132643.279$-$472816.18		&		0.43		&	0.32	&		15/14.6 		&		9/8.3   		&		8/8.3   		&		0.00		&		6.3		&		9.1		&		3.6		&		\ldots		&	 \ldots \\
{\bf 11f}		&		J132642.670$-$472723.56		&		0.49		&	0.61	&		12/11.7 		&		7/6.7   		&		5/4.5   		&		0.17		&		5.0		&		6.3		&		4.1		&		\ldots		&	 \ldots \\
12a			&		J132648.643$-$472744.46		&		0.36		&	0.41	&		35/36.8 		&		23/22.6 		&		12/13.3 		&		0.23		&		16.0		&		20.1		&		10.4		&		0.07		&	  CV \\
12b			&		J132652.514$-$472737.70		&		0.33		&	0.56	&		116/133.9		&		56/60.0 		&		69/91.9 		&		$-0.19$		&		56.4		&		79.0		&		32.7		&		0.48		&	 \ldots \\
{\bf 12c}		&		J132647.945$-$472810.74		&		0.41		&	0.23	&		16/15.5 		&		10/9.4  		&		6/5.6   		&		0.22		&		6.6		&		8.6		&		4.0		&		\ldots		&	 \ldots \\
13a			&		J132653.506$-$472900.17		&		0.29		&	0.42	&		1860/2140.1		&		589/622.3		&		1464/1955.6		&		$-0.50$		&		914.0		&		1311.3		&		429.7		&		1.17		&	  CV \\
13b			&		J132650.563$-$472918.19		&		0.33		&	0.30	&		17/92.8 		&		5/24.2  		&		14/87.4 		&		$-0.56$		&		34.2		&		48.4		&		16.8		&		1.43		&	 \ldots \\ \thispagestyle{empty}
13c			&		J132652.126$-$472935.45		&		0.29		&	0.45	&		2042/3004.5		&		782/1047.9		&		1470/2471.0		&		$-0.37$		&		1238.5		&		1774.5		&		638.1		&		1.68		&	 CV \\
13d			&		J132649.574$-$472924.18		&		0.36		&	0.29	&		6/34.4  		&		6/32.9  		&		0/0.0   		&		\ldots		&		16.0		&		20.3		&		13.9		&		0.97		&	 \ldots \\
13e			&		J132646.246$-$472948.27		&		0.42		&	0.40	&		15/17.0   		&		13/14.3 		&		2/1.1   		&		1.11		&		6.9		&		8.8		&		5.3		&		0.65		&	 \ldots \\
13f			&		J132645.984$-$472916.32		&		0.34		&	0.21	&		17/67.9 		&		6/21.5  		&		14/64.5 		&		$-0.48$		&		38.7		&		58.8		&		21.3		&		3.08		&	 CV? \\
{\bf 13g}		&		J132654.065$-$472859.06		&		0.41		&	0.45	&		16/16.9 		&		9/9.4   		&		7/7.2   		&		0.12		&		7.4		&		9.1		&		4.8		&		\ldots		&	 \ldots \\
14c			&		J132644.062$-$472856.47		&		0.36		&	0.22	&		44/45.9 		&		41/39.9 		&		4/3.7   		&		1.03		&		19.7		&		26.0		&		16.0		&		0.58		&	 \ldots \\
14d			&		J132638.016$-$472910.48		&		0.39		&	0.62	&		16/45.2 		&		3/7.4   		&		17/55.9 		&		$-0.88$		&		21.2		&		34.3		&		9.0		&		1.36		&	 \ldots \\
{\bf 14e}		&		J132641.858$-$472923.03		&		0.55		&	0.42	&		7/7.5   		&		2/1.8   		&		5/6.2   		&		$-0.54$		&		3.3		&		4.1		&		0.7		&		\ldots		&	 \ldots \\
{\bf 14f}		&		J132639.154$-$472842.79		&		0.43		&	0.53	&		22/22.0 		&		5/4.4   		&		18/20.6 		&		$-0.67$		&		9.5		&		12.6		&		4.4		&		\ldots		&	 \ldots \\
21a			&		J132631.183$-$472827.57		&		0.43		&	1.06	&		58/62.0 		&		15/14.5 		&		44/54.5 		&		$-0.57$		&		26.9		&		35.6		&		13.7		&		1.02		&	 \ldots \\
21c			&		J132636.852$-$472745.84		&		0.56		&	0.78	&		12/11.6 		&		5/4.6   		&		9/9.3   		&		$-0.31$		&		5.0		&		7.3		&		2.5		&		0.80		&	 \ldots \\
21d			&		J132638.258$-$472740.12		&		0.38		&	0.73	&		59/62.4 		&		27/26.5 		&		35/42.0   		&		$-0.20$		&		27.1		&		36.4		&		16.0		&		1.03		&	 \ldots \\
21e			&		J132645.206$-$472652.25		&		0.39		&	0.75	&		38/42.3 		&		31/32.3 		&		7/7.6   		&		0.63		&		17.7		&		22.6		&		13.9		&		0.96		&	 \ldots \\
{\bf 21f}		&		J132637.860$-$472715.51		&		0.48		&	0.85	&		20/20.6 		&		16/15.7 		&		4/3.6   		&		0.64		&		8.9		&		11.3		&		6.9		&		\ldots		&	 \ldots \\
{\bf 21g}		&		J132640.776$-$472700.84		&		0.58		&	0.80	&		10/9.4  		&		2/1.5   		&		10/10.9 		&		$-0.86$		&		4.0		&		6.2		&		1.3		&		\ldots		&	 \ldots \\
22a			&		J132648.259$-$472640.76		&		0.52		&	0.81	&		12/11.8 		&		7/6.5   		&		5/5.1   		&		0.11		&		5.2		&		6.5		&		2.5		&		0.76		&	 (FGND) \\
22c			&		J132652.682$-$472713.10		&		0.32		&	0.70	&		179/212.6		&		96/105.8		&		97/133.4		&		$-0.10$		&		88.5		&		123.1		&		55.3		&		1.23		&	 fbCV?\\
22d			&		J132658.723$-$472728.73		&		0.44		&	0.90	&		20/21.9 		&		14/14.6 		&		6/6.7   		&		0.34		&		9.5		&		12.2		&		7.3		&		0.47		&	 \ldots \\
22e			&		J132659.930$-$472809.50		&		0.39		&	0.86	&		27/39.3 		&		10/13.6 		&		19/31.7 		&		$-0.37$		&		17.8		&		24.4		&		12.5		&		0.45		&	 RGB/SGB-a \\ \thispagestyle{empty}
22f			&		J132658.802$-$472820.90		&		0.36		&	0.77	&		51/68.3 		&		23/28.7 		&		29/44.3 		&		$-0.19$		&		28.2		&		37.2		&		15.2		&		1.18		&	 (AGN) \\
{\bf 22g}		&		J132646.181$-$472640.95		&		0.72		&	0.81	&		6/5.2   		&		0/0.0   		&		9/9.7   		&		\ldots		&		2.2		&		4.4		&		0.0		&		\ldots		&	 \ldots \\
{\bf 22h}		&		J132657.638$-$472712.29		&		0.49		&	0.91	&		14/14.4 		&		6/5.7   		&		10/11.8 		&		$-0.31$		&		6.1		&		9.1		&		3.4		&		\ldots		&	 \ldots \\
{\bf 22i}		&		J132656.290$-$472806.82		&		0.53		&	0.65	&		6/7.2   		&		5/6.2   		&		1/0.1   		&		1.96		&		3.5		&		4.3		&		2.6		&		\ldots		&	 \ldots \\
{\bf 22j}		&		J132656.460$-$472815.67		&		0.46		&	0.64	&		10/11.3 		&		8/8.9   		&		2/1.6   		&		0.76		&		5.0		&		6.1		&		3.8		&		\ldots		&	 \ldots \\
{\bf 22k}		&		J132659.328$-$472829.08		&		0.65		&	0.80	&		6/5.3   		&		1/0.6   		&		8/8.9   		&		$-1.16$		&		2.2		&		4.5		&		0.3		&		\ldots		&	 \ldots \\
{\bf 22l}		&		J132654.828$-$472832.66		&		0.40		&	0.51	&		20/19.6 		&		13/12.3 		&		9/8.9   		&		0.14		&		8.4		&		11.6		&		6.2		&		\ldots		&	 \ldots \\
23a			&		J132651.050$-$473009.98		&		0.36		&	0.59	&		48/58.6 		&		21/23.6 		&		31/42.2 		&		$-0.25$		&		24.3		&		33.6		&		15.3		&		0.54		&	 \ldots \\
23b			&		J132651.672$-$473047.21		&		0.47		&	0.83	&		19/18.5 		&		5/4.4   		&		14/15.2 		&		$-0.54$		&		7.9		&		10.0		&		5.2		&		0.57		&	 CV? \\
{\bf 23d}		&		J132658.476$-$472847.45		&		0.47		&	0.74	&		13/12.3 		&		6/5.3   		&		10/11.0   		&		$-0.32$		&		5.2		&		8.4		&		3.1		&		\ldots		&	 \ldots \\
{\bf 23e}		&		J132658.174$-$472957.10		&		0.55		&	0.85	&		9/9.8   		&		7/7.6   		&		4/3.9   		&		0.29		&		4.2		&		6.5		&		3.3		&		\ldots		&	 \ldots \\
{\bf 23f}		&		J132657.247$-$473012.74		&		0.55		&	0.86	&		9/10.3  		&		6/6.5   		&		3/3.1   		&		0.33		&		4.6		&		5.7		&		3.0		&		\ldots		&	 \ldots \\
{\bf 23g}		&		J132655.231$-$473011.63		&		0.46		&	0.76	&		14/15.8 		&		12/13.2 		&		3/2.8   		&		0.68		&		6.7		&		9.2		&		5.5		&		\ldots		&	 \ldots \\
{\bf 23h}		&		J132653.945$-$473036.11		&		0.55		&	0.83	&		11/10.5 		&		9/8.5   		&		2/0.9   		&		0.97		&		4.4		&		5.5		&		3.6		&		\ldots		&	 \ldots \\
24b			&		J132639.312$-$473037.06		&		0.50		&	0.88	&		20/20.0 		&		14/13.4 		&		6/5.6   		&		0.38		&		8.6		&		10.9		&		6.0		&		1.02		&	 \ldots \\
24c			&		J132638.422$-$473036.53		&		0.40		&	0.92	&		71/76.1 		&		34/33.8 		&		41/49.7 		&		$-0.17$		&		32.3		&		44.1		&		19.8		&		1.23		&	 CV \\
24d			&		J132637.399$-$473006.97		&		0.49		&	0.83	&		20/21.2 		&		14/14.3 		&		6/6.1   		&		0.37		&		9.2		&		11.9		&		6.4		&		0.98		&	 \ldots \\
24f			&		J132637.248$-$472942.37		&		0.45		&	0.75	&		20/26.0 		&		17/21.2 		&		8/10.8  		&		0.29		&		11.1		&		17.8		&		8.6		&		1.72		&	 RGB/SGB-a \\
24g			&		J132634.394$-$472955.46		&		0.39		&	0.95	&		82/93.7 		&		41/43.2 		&		50/65.2 		&		$-0.18$		&		40.9		&		59.1		&		21.4		&		0.78		&	 (AGN?) \\ \thispagestyle{empty}
{\bf 24i}		&		J132642.564$-$473022.49		&		0.49		&	0.69	&		16/15.8 		&		1/0.4   		&		20/23.1 		&		$-1.71$		&		6.7		&		11.3		&		1.3		&		\ldots		&	 \ldots \\
{\bf 24j}		&		J132642.326$-$473019.33		&		0.45		&	0.68	&		21/20.9 		&		16/15.0 		&		5/4.7   		&		0.50		&		8.9		&		11.3		&		7.4		&		\ldots		&	 \ldots \\
{\bf 24k}		&		J132635.422$-$472838.24		&		0.81		&	0.77	&		6/5.1   		&		2/1.4   		&		8/8.6   		&		$-0.78$		&		2.4		&		5.3		&		1.0		&		\ldots		&	 \ldots \\
31a			&		J132629.359$-$472813.08		&		0.40		&	1.19	&		119/136.9		&		51/54.1 		&		79/105.7		&		$-0.29$		&		58.8		&		84.0		&		33.5		&		1.58		&	 fbCV? \\
31b			&		J132631.414$-$472801.23		&		0.56		&	1.08	&		20/20.2 		&		8/7.4   		&		12/13.7 		&		$-0.27$		&		8.9		&		11.3		&		5.7		&		1.01		&	 (AGN) \\
31c			&		J132632.323$-$472707.93		&		0.60		&	1.16	&		17/18.3 		&		12/12.4 		&		5/4.9   		&		0.40		&		7.9		&		9.8		&		6.9		&		0.45		&	 \ldots \\
31d			&		J132636.230$-$472621.77		&		0.42		&	1.18	&		74/83.6 		&		32/33.6 		&		46/59.9 		&		$-0.25$		&		36.2		&		49.1		&		19.8		&		0.75		&	 \ldots \\
{\bf 31e}		&		J132630.415$-$472711.19		&		0.72		&	1.26	&		12/13.3 		&		7/7.5   		&		5/5.3   		&		0.15		&		5.7		&		7.1		&		3.3		&		\ldots		&	 \ldots \\
{\bf 31f}		&		J132630.470$-$472618.39		&		0.62		&	1.45	&		21/26.6 		&		3/3.3   		&		20/29.8 		&		$-0.95$		&		11.3		&		16.2		&		1.8		&		\ldots		&	 \ldots \\
{\bf 31g}		&		J132634.171$-$472641.86		&		0.63		&	1.17	&		14/15.4 		&		9/9.6   		&		6/6.8   		&		0.15		&		6.5		&		8.9		&		5.0		&		\ldots		&	 \ldots \\
32a			&		J132646.356$-$472518.21		&		0.55		&	1.35	&		22/25.6 		&		6/6.2   		&		17/23.1 		&		$-0.57$		&		10.9		&		14.8		&		5.9		&		0.75		&	 fbCV? \\
32b			&		J132652.162$-$472532.32		&		0.55		&	1.29	&		8/21.2  		&		3/7.2   		&		5/15.0  		&		$-0.32$		&		8.8		&		11.3		&		4.7		&		1.03		&	 \ldots \\
32c			&		J132655.894$-$472602.06		&		0.44		&	1.20	&		38/40.9 		&		10/9.8  		&		34/42.8 		&		$-0.64$		&		17.4		&		26.3		&		9.0		&		1.68		&	 \ldots \\
32d			&		J132702.422$-$472647.03		&		0.55		&	1.26	&		17/17.6 		&		13/13.1 		&		5/4.8   		&		0.44		&		7.5		&		10.1		&		5.4		&		0.61		&	 \ldots \\
32f			&		J132705.328$-$472808.56		&		0.39		&	1.21	&		73/84.8 		&		41/44.5 		&		35/46.6 		&		$-0.02$		&		36.0		&		48.7		&		25.1		&		1.55		&	 RGB/SGB-a \\
{\bf 32g}		&		J132647.410$-$472549.21		&		0.71		&	1.14	&		9/8.8   		&		5/4.7   		&		4/4.0   		&		0.08		&		3.7		&		4.8		&		3.0		&		\ldots		&	 \ldots \\
{\bf 32h}		&		J132650.686$-$472512.94		&		0.48		&	1.40	&		36/47.0   		&		20/24.1 		&		17/25.1 		&		$-0.02$		&		20.0		&		26.6		&		13.3		&		\ldots		&	 \ldots \\
{\bf 32i}		&		J132658.752$-$472713.88		&		0.83		&	0.96	&		5/4.1   		&		3/2.6   		&		3/2.5   		&		0.01		&		1.7		&		2.8		&		1.3		&		\ldots		&	 \ldots \\
33b			&		J132706.478$-$472852.93		&		0.45		&	1.26	&		39/41.2 		&		7/6.6   		&		37/45.7 		&		$-0.84$		&		17.4		&		25.6		&		7.5		&		1.08		&	 \ldots \\
33c			&		J132703.564$-$472857.59		&		0.65		&	1.07	&		9/8.0   		&		4/3.3   		&		6/6.0   		&		$-0.25$		&		3.3		&		4.8		&		1.9		&		0.49		&	 \ldots \\
33d			&		J132701.500$-$472924.60		&		0.39		&	0.97	&		24/56.5 		&		6/12.7  		&		21/57.6 		&		$-0.66$		&		22.9		&		33.3		&		13.7		&		0.96		&	 \ldots \\ \thispagestyle{empty}
33e			&		J132700.970$-$473004.42		&		0.43		&	1.03	&		28/32.8 		&		14/15.4 		&		16/20.7 		&		$-0.13$		&		13.8		&		18.9		&		10.2		&		0.94		&	 CV? \\
33f			&		J132659.280$-$473037.45		&		0.51		&	1.07	&		19/19.4 		&		7/6.4   		&		13/15.0 		&		$-0.37$		&		8.1		&		10.9		&		4.4		&		0.65		&	 \ldots \\
33g			&		J132655.997$-$473045.87		&		0.59		&	0.96	&		11/10.3 		&		10/9.5  		&		1/0.0   		&		\ldots		&		4.4		&		5.5		&		3.7		&		0.60		&	 \ldots \\
33h			&		J132655.058$-$473113.25		&		0.35		&	1.08	&		247/271.2		&		93/93.6 		&		174/218.5		&		$-0.37$		&		114.9		&		160.6		&		62.3		&		1.53		&	 \ldots \\
33i			&		J132651.070$-$473144.34		&		0.40		&	1.18	&		96/106.3		&		36/36.5  		&		70/88.8 		&		$-0.39$		&		45.2		&		64.7		&		25.1		&		0.49		&	 \ldots \\
33k			&		J132649.788$-$473147.66		&		0.41		&	1.18	&		65/87.7 		&		30/37.2 		&		40/61.4 		&		$-0.22$		&		37.0		&		51.9		&		22.3		&		1.24		&	 \ldots \\
33l			&		J132648.725$-$473124.90		&		0.32		&	1.03	&		307/897.5		&		154/411.1		&		172/577.5		&		$-0.15$		&		366.5		&		505.1		&		221.2		&		2.74		&	 (AGN?) \\
33m			&		J132646.500$-$473140.77		&		0.49		&	1.13	&		28/32.2 		&		8/8.3   		&		23/30.5 		&		$-0.56$		&		13.5		&		19.4		&		6.2		&		1.17		&	 CV? \\
{\bf 33n}		&		J132703.175$-$472901.10		&		0.90		&	1.05	&		5/4.0   		&		2/1.5   		&		4/3.8   		&		$-0.40$		&		1.7		&		2.7		&		0.9		&		\ldots		&	 \ldots \\
{\bf 33o}		&		J132700.106$-$473036.29		&		0.55		&	1.10	&		15/15.1 		&		4/3.6   		&		13/15.0 		&		$-0.62$		&		6.4		&		9.3		&		2.1		&		\ldots		&	 \ldots \\
{\bf 33p}		&		J132657.691$-$473021.51		&		0.68		&	0.92	&		6/6.5   		&		2/1.8   		&		4/4.5   		&		$-0.40$		&		2.8		&		3.4		&		1.6		&		\ldots		&	 \ldots \\
{\bf 33q}		&		J132652.366$-$473153.45		&		0.76		&	1.25	&		10/9.9  		&		4/3.6   		&		6/6.5   		&		$-0.25$		&		4.4		&		5.5		&		3.0		&		\ldots		&	 \ldots \\
34b			&		J132637.421$-$473052.75		&		0.39		&	1.04	&		87/103.5		&		30/32.6 		&		63/86.4 		&		$-0.42$		&		46.2		&		63.8		&		25.6		&		0.32		&	 RGB/SGB-a \\
34d			&		J132634.330$-$473032.98		&		0.43		&	1.09	&		56/64.1 		&		16/16.8 		&		47/62.2 		&		$-0.57$		&		26.6		&		39.2		&		13.0		&		0.35		&	 \ldots \\
{\bf 34e}		&		J132643.202$-$473110.62		&		0.51		&	0.97	&		21/21.0 		&		12/11.3 		&		9/9.3   		&		0.09		&		8.9		&		11.3		&		6.0		&		\ldots		&	 \ldots \\
{\bf 34f}		&		J132631.711$-$473105.93		&		0.85		&	1.36	&		10/11.5 		&		5/5.5   		&		5/5.9   		&		$-0.03$		&		4.9		&		6.1		&		2.6		&		\ldots		&	 \ldots \\
{\bf 34g}		&		J132631.877$-$473101.63		&		0.78		&	1.33	&		12/13.4 		&		2/1.8   		&		12/15.9 		&		$-0.95$		&		5.6		&		8.6		&		1.5		&		\ldots		&	 \ldots \\
{\bf 34h}		&		J132631.344$-$472932.22		&		0.61		&	1.08	&		15/16.8 		&		9/9.6   		&		7/8.3   		&		0.06		&		7.0		&		9.6		&		4.5		&		\ldots		&	 \ldots \\
41a			&		J132624.398$-$472657.40		&		0.56		&	1.65	&		50/56.8 		&		25/26.9 		&		28/36.1 		&		$-0.13$		&		24.7		&		34.2		&		13.6		&		1.27		&	 \ldots \\
41c			&		J132624.535$-$472610.61		&		0.62		&	1.79	&		44/52.6 		&		15/16.5 		&		30/41.4 		&		$-0.40$		&		22.8		&		30.2		&		10.2		&		1.09		&	 \ldots \\ \thispagestyle{empty}
41d			&		J132628.644$-$472627.04		&		0.38		&	1.51	&		360/407.5		&		73/76.5 		&		334/438.9		&		$-0.76$		&		177.7		&		262.5		&		65.7		&		1.17		&	 CV \\
41e			&		J132630.547$-$472600.81		&		0.60		&	1.53	&		34/34.7 		&		28/28.3 		&		7/5.4   		&		0.72		&		15.1		&		19.9		&		12.3		&		1.06		&	 \ldots \\
41f			&		J132632.016$-$472451.22		&		0.48		&	1.82	&		102/140.2		&		52/67.5 		&		56/88.3 		&		$-0.12$		&		59.0		&		82.8		&		36.7		&		1.44		&	 \ldots \\
41g			&		J132637.339$-$472430.15		&		0.57		&	1.78	&		41/58.4 		&		22/29.9 		&		23/36.8 		&		$-0.09$		&		25.5		&		36.9		&		17.6		&		0.73		&	 (AGN) \\
{\bf 41i}		&		J132623.921$-$472731.35		&		0.91		&	1.60	&		15/14.4 		&		4/3.4    		&		11/11.6 		&		$-0.53$		&		6.1		&		7.8		&		3.3		&		\ldots		&	 \ldots \\
{\bf 41j}		&		J132624.307$-$472646.93		&		0.84		&	1.69	&		18/19.4 		&		6/6.3   		&		13/15.0 		&		$-0.38$		&		8.6		&		11.4		&		5.1		&		\ldots		&	 \ldots \\
{\bf 41k}		&		J132630.958$-$472521.71		&		0.66		&	1.70	&		30/32.7 		&		19/20.7 		&		13/14.5 		&		0.15		&		14.3		&		19.7		&		9.8		&		\ldots		&	 \ldots \\
{\bf 41l}		&		J132636.413$-$472528.70		&		0.76		&	1.46	&		12/14.7 		&		7/8.2   		&		7/9.7   		&		$-0.07$		&		6.1		&		9.6		&		4.2		&		\ldots		&	 \ldots \\
42a			&		J132701.651$-$472543.35		&		0.72		&	1.51	&		11/13.9 		&		5/5.9   		&		7/9.9   		&		$-0.23$		&		5.7		&		8.1		&		4.7		&		0.34		&	 \ldots \\
{\bf 42d}		&		J132649.128$-$472450.90		&		0.61		&	1.53	&		17/25.1 		&		14/19.6 		&		3/4.0   		&		0.69		&		10.4		&		14.0		&		9.0		&		\ldots		&	 \ldots \\
43a			&		J132707.956$-$472944.51		&		0.46		&	1.41	&		17/50.2 		&		9/24.7  		&		8/26.2  		&		$-0.03$		&		21.2		&		26.8		&		13.6		&		0.84		&	 \ldots \\
43c			&		J132706.890$-$473008.85		&		0.53		&	1.39	&		22/27.0   		&		10/11.4 		&		12/16.3 		&		$-0.16$		&		11.2		&		14.4		&		7.0		&		1.00		&	 RGB/SGB-a \\
43d			&		J132704.495$-$473037.64		&		0.42		&	1.34	&		72/80.7 		&		27/28.0 		&		50/63.9  		&		$-0.36$		&		33.7		&		47.1		&		17.6		&		1.16		&	 \ldots \\
43f			&		J132656.014$-$473202.05		&		0.61		&	1.39	&		19/21.3 		&		14/14.9 		&		5/5.2   		&		0.45		&		8.7		&		11.4		&		7.0		&		0.39		&	 (FGND) \\
43h			&		J132649.582$-$473212.70		&		0.37		&	1.34	&		208/290.8		&		96/122.7		&		128/207.2		&		$-0.23$		&		121.4		&		171.7		&		68.7		&		1.75		&	 CV? \\
{\bf 43i}		&		J132704.666$-$473008.55		&		0.69		&	1.26	&		9/10.0  		&		0/0.0   		&		12/15.6 		&		\ldots		&		4.2		&		7.2		&		0.0		&		\ldots		&	 \ldots \\
{\bf 43j}		&		J132705.256$-$473015.81		&		0.57		&	1.31	&		13/18.9 		&		8/11.2  		&		6/9.0   		&		0.09		&		7.8		&		10.8		&		5.8		&		\ldots		&	 \ldots \\
{\bf 43k}		&		J132701.380$-$473059.64		&		0.61		&	1.26	&		15/15.1 		&		4/3.5   		&		17/20.3 		&		$-0.76$		&		6.3		&		11.8		&		2.9		&		\ldots		&	 \ldots \\
{\bf 43l}		&		J132657.187$-$473134.17		&		0.66		&	1.26	&		13/13.5 		&		2/1.7   		&		12/14.3 		&		$-0.92$		&		5.6		&		7.9		&		1.5		&		\ldots		&	 \ldots \\
{\bf 43m}		&		J132652.704$-$473210.42		&		0.91		&	1.36	&		8/8.4   		&		3/2.8   		&		6/7.1   		&		$-0.40$		&		3.3		&		5.1		&		1.5		&		\ldots		&	 \ldots \\ \thispagestyle{empty}
44a			&		J132644.110$-$473230.86		&		0.39		&	1.46	&		176/271.0 		&		105/148.4		&		79/142.4		&		0.02		&		112.0		&		156.2		&		76.9		&		0.51		&	 \ldots \\
44b			&		J132633.468$-$473153.25		&		0.64		&	1.50	&		26/29.1 		&		4/3.4   		&		27/35.1 		&		$-1.01$		&		12.6		&		19.2		&		3.9		&		0.75		&	 \ldots \\
44c			&		J132623.674$-$473044.56		&		0.94		&	1.72	&		16/16.9 		&		2/1.7   		&		15/18.2 		&		$-1.03$		&		7.4		&		9.9		&		3.2		&		0.41		&	 fbCV? \\
44d			&		J132622.886$-$473008.58		&		0.65		&	1.68	&		36/39.2 		&		16/16.4 		&		26/32.1 		&		$-0.29$		&		16.5		&		25.8		&		10.4		&		0.56		&	 CV? \\
44e			&		J132619.793$-$472910.32		&		0.38		&	1.80	&		882/1066.4		&		754/859.2		&		129/174.8		&		0.69		&		463.3		&		600.0		&		385.9		&		0.71$^g$ 	&	 qLMXB \\
{\bf 44f}		&		J132626.616$-$473025.17		&		1.01		&	1.49	&		12/10.5 		&		2/1.2   		&		12/12.7 		&		$-1.02$		&		4.3		&		6.7		&		1.0		&		\ldots		&	 \ldots \\
51a			&		J132631.286$-$472439.00		&		0.56		&	1.91	&		65/82.0 		&		26/31.5 		&		46/64.6 		&		$-0.31$		&		36.6		&		51.8		&		17.9		&		0.69		&	 \ldots \\
51b			&		J132630.614$-$472346.31		&		1.21		&	2.22	&		18/18.2 		&		5/5.0   		&		15/15.7 		&		$-0.50$		&		7.7		&		10.5		&		4.5		&		0.59		&	 \ldots \\
51d			&		J132640.980$-$472401.75		&		0.47		&	1.88	&		130/164.4		&		52/62.3 		&		81/118.4		&		$-0.28$		&		70.7		&		95.3		&		45.9		&		1.90		&	 \ldots \\
51e			&		J132644.724$-$472333.32		&		0.94		&	2.03	&		22/20.4 		&		14/14.4 		&		8/4.6   		&		0.50		&		8.8		&		10.7		&		7.2		&		0.62		&	 (FGND) \\
{\bf 51f}		&		J132629.460$-$472403.01		&		1.18		&	2.17	&		19/18.0 		&		3/2.1   		&		24/28.9 		&		$-1.14$		&		7.6		&		15.2		&		2.4		&		\ldots		&	 \ldots \\
52c			&		J132706.394$-$472537.90		&		0.46		&	1.75	&		93/105.0		&		42/44.2 		&		63/83.0 		&		$-0.27$		&		45.0		&		66.0		&		22.5		&		0.79		&	 (AGN?) \\
52d			&		J132714.993$-$472743.49		&		0.68		&	1.86	&		24/27.4 		&		10/10.9 		&		18/23.6 		&		$-0.33$		&		12.0		&		18.1		&		7.0		&		0.84		&	 fbCV? \\
52e			&		J132717.066$-$472819.07		&		0.47		&	1.96	&		114/143.5		&		45/53.0 		&		81/120.1		&		$-0.36$		&		60.3		&		88.0		&		32.6		&		4.51		&	 \ldots \\
52f			&		J132714.395$-$472830.55		&		0.57		&	1.78	&		34/43.3 		&		22/26.9 		&		12/16.7 		&		0.21		&		18.2		&		23.8		&		12.1		&		0.80		&	 \ldots \\
{\bf 52g}		&		J132717.698$-$472748.52		&		0.74		&	2.03	&		24/29.2 		&		15/18.1 		&		9/10.6  		&		0.23		&		12.5		&		15.5		&		9.5		&		\ldots		&	 \ldots \\
53a			&		J132659.767$-$473237.12		&		0.74		&	1.70	&		19/21.2 		&		14/15.7 		&		7/6.7   		&		0.37		&		9.0		&		12.7		&		6.4		&		0.45		&	 \ldots \\
{\bf 53c}		&		J132701.651$-$473259.99		&		1.04		&	1.89	&		13/13.6 		&		6/6.4   		&		7/6.4   		&		0.00		&		5.7		&		6.9		&		2.7		&		\ldots		&	 \ldots \\
{\bf 53d}		&		J132700.163$-$473303.61		&		0.91		&	1.86	&		15/17.3 		&		9/10.6  		&		6/5.3   		&		0.31		&		7.0		&		8.6		&		5.2		&		\ldots		&	 \ldots \\
54b			&		J132642.461$-$473308.93		&		0.48		&	1.72	&		103/121.5		&		13/13.7 		&		112/153.6		&		$-1.05$		&		52.5		&		83.0		&		18.0		&		1.60		&	 fbCV? \\ \thispagestyle{empty}
54c			&		J132627.050$-$473215.02		&		0.48		&	1.89	&		153/195.2		&		31/36.7 		&		137/202.2		&		$-0.74$		&		88.1		&		124.9		&		37.5		&		2.89		&	 \ldots \\
54d			&		J132625.128$-$473227.15		&		0.46		&	2.03	&		224/346.1		&		107/155.2		&		137/245.9		&		$-0.20$		&		156.1		&		221.5		&		99.2		&		0.71		&	 (AGN) \\
54f			&		J132621.898$-$473049.42		&		0.86		&	1.84	&		26/24.7 		&		18/17.9 		&		11/7.8  		&		0.36		&		10.7		&		14.7		&		7.3		&		0.98		&	 \ldots \\
54g			&		J132620.138$-$473015.69		&		0.87		&	1.86	&		29/25.0 		&		12/10.0 		&		18/16.2 		&		$-0.21$		&		10.8		&		14.0		&		6.7		&		2.14		&	 (AGN) \\
54h			&		J132620.362$-$473002.68		&		0.42		&	1.83	&		394/434.3		&		139/142.7		&		301/384.2		&		$-0.43$		&		191.1		&		276.5		&		97.3		&		0.54		&	 CV \\
{\bf 54i}		&		J132623.978$-$473159.12		&		1.48		&	1.96	&		15/10.7 		&		4/1.8   		&		15/13.9 		&		$-0.89$		&		4.5		&		7.8		&		1.3		&		\ldots		&	 \ldots \\
{\bf 54j}		&		J132622.716$-$473212.60		&		1.15		&	2.08	&		19/19.3 		&		5/3.5   		&		16/20.7 		&		$-0.77$		&		8.2		&		12.0		&		2.1		&		\ldots		&	 \ldots \\
{\bf 54k}		&		J132617.566$-$473052.63		&		1.11		&	2.10	&		22/21.4 		&		8/7.3   		&		14/13.3 		&		$-0.26$		&		8.9		&		10.8		&		5.0		&		\ldots		&	 \ldots \\
{\bf 54l}		&		J132616.634$-$473053.66		&		0.62		&	2.16	&		66/113.6		&		24/39.2 		&		51/103.6		&		$-0.42$		&		48.5		&		74.9		&		24.6		&		\ldots		&	 \ldots \\
{\bf 54m}		&		J132615.161$-$473021.83		&		1.19		&	2.19	&		27/21.0 		&		16/14.0 		&		15/10.7 		&		0.12		&		9.4		&		13.8		&		6.1		&		\ldots		&	 \ldots \\
{\bf 54n}		&		J132617.266$-$472951.50		&		1.26		&	2.01	&		19/15.0 		&		15/14.3 		&		5/0.0   		&		\ldots		&		6.6		&		8.5		&		5.2		&		\ldots		&	 \ldots \\
61a			&		J132620.930$-$472448.22		&		0.97		&	2.31	&		33/34.2 		&		8/7.2   		&		34/42.4 		&		$-0.77$		&		14.9		&		25.2		&		4.2		&		3.32		&	 \ldots \\
61b			&		J132641.510$-$472216.08		&		0.52		&	2.55	&		274/388.7		&		74/107.1		&		234/383.4		&		$-0.55$		&		173.7		&		260.5		&		82.0		&		1.03		&	 \ldots \\
{\bf 61c}		&		J132612.600$-$472831.80		&		1.08		&	2.27	&		20/28.3 		&		6/2.4   		&		17/35.4 		&		$-1.16$		&		11.4		&		18.5		&		3.2		&		\ldots		&	 \ldots \\
{\bf 61d}		&		J132610.219$-$472657.68		&		1.27		&	2.52	&		30/28.1 		&		9/7.2   		&		26/28.9 		&		$-0.60$		&		12.4		&		18.9		&		5.8		&		\ldots		&	 \ldots \\
62a			&		J132648.437$-$472216.95		&		0.70		&	2.52	&		69/92.1 		&		27/38.3 		&		51/74.4 		&		$-0.29$		&		41.3		&		59.1		&		20.6		&		2.92		&	 \ldots \\
62b			&		J132707.990$-$472333.64		&		0.51		&	2.43	&		195/270.7		&		64/87.4 		&		141/229.4		&		$-0.42$		&		112.6		&		159.4		&		58.7		&		1.45		&	 (AGN) \\
62c			&		J132712.415$-$472423.95		&		0.68		&	2.36	&		52/66.8 		&		22/28.0 		&		36/53.0 		&		$-0.28$		&		28.6		&		40.9		&		14.9		&		1.03		&	 \ldots \\
62d			&		J132716.253$-$472457.73		&		0.80		&	2.41	&		24/45.5 		&		20/39.6 		&		4/0.6   		&		1.83		&		18.1		&		22.7		&		14.4		&		0.93		&	 \ldots \\
62e			&		J132723.978$-$472819.07		&		0.96		&	2.41	&		24/28.9 		&		7/8.1   		&		21/29.2 		&		$-0.56$		&		11.9		&		18.4		&		7.1		&		0.68		&	 \ldots \\ \thispagestyle{empty}
{\bf 62f}		&		J132706.386$-$472315.61		&		1.06		&	2.48	&		28/26.9 		&		4/2.5   		&		28/32.6 		&		$-1.12$		&		11.4		&		17.3		&		5.4		&		\ldots		&	 \ldots \\
{\bf 62g}		&		J132707.488$-$472345.08		&		1.01		&	2.35	&		23/24.9 		&		12/13.8 		&		12/12.1 		&		0.06		&		10.6		&		13.6		&		7.6		&		\ldots		&	 \ldots \\
{\bf 62h}		&		J132719.939$-$472550.10		&		1.12		&	2.42	&		20/21.3 		&		6/5.5   		&		16/20.1 		&		$-0.56$		&		8.7		&		12.5		&		4.7		&		\ldots		&	 \ldots \\
63a			&		J132722.918$-$472907.38		&		1.06		&	2.34	&		22/21.4         	&		5/4.4   		&		23/26.7 		&		$-0.78$		&		9.0		&		15.0		&		3.7		&		0.33		&	 \ldots \\
63b			&		J132714.623$-$473150.71		&		0.55		&	2.15	&		99/115.8		&		16/17.0   		&		97/130.7		&		$-0.89$		&		48.8		&		71.5		&		18.0		&		1.51		&	 \ldots \\
63c			&		J132711.750$-$473240.79		&		0.51		&	2.20	&		153/178.4		&		61/67.0 		&		106/141.0 		&		$-0.32$		&		74.2		&		105.4		&		41.4		&		0.22		&	 \ldots \\
63d			&		J132710.030$-$473320.24		&		0.54		&	2.31	&		126/177.0		&		28/36.1 		&		123/202.8		&		$-0.75$		&		73.1		&		115.7		&		27.4		&		0.70		&	 \ldots \\
63e			&		J132701.932$-$473355.31		&		1.07		&	2.21	&		18/21.3 		&		3/2.5   		&		19/26.9 		&		$-1.03$		&		9.3		&		14.4		&		0.8		&		0.55		&	 \ldots \\
63f			&		J132657.175$-$473351.85		&		0.99		&	2.08	&		20/21.0 		&		18/21.5 		&		4/0.0   		&		\ldots		&		8.9		&		12.3		&		8.1		&		0.64		&	 \ldots \\
63g			&		J132656.683$-$473427.68		&		0.76		&	2.29	&		47/56.1 		&		4/3.1   		&		49/68.8 		&		$-1.34$		&		23.6		&		34.4		&		11.6		&		1.26		&	 \ldots \\
{\bf 63h}		&		J132721.562$-$472920.91		&		1.04		&	2.26	&		21/19.6 		&		18/19.3 		&		3/0.0   		&		\ldots		&		8.3		&		9.5		&		7.2		&		\ldots		&	 (NV410) \\
{\bf 63i}		&		J132712.418$-$473251.55		&		1.03		&	2.28	&		22/23.5 		&		1/0.0   		&		24/31.4 		&		\ldots		&		9.7		&		13.9		&		0.1		&		\ldots		&	 \ldots \\
64a			&		J132627.012$-$473407.75		&		0.69		&	2.46	&		100/119.8		&		25/28.1 		&		86/119.4		&		$-0.63$		&		51.6		&		75.4		&		23.6		&		0.56		&	 \ldots \\
64b			&		J132624.437$-$473302.58		&		0.90		&	2.23	&		39/39.1 		&		2/0.0   		&		51/64.6 		&		\ldots		&		17.3		&		31.3		&		2.2		&		0.68		&	 \ldots \\
64c			&		J132616.279$-$473058.39		&		0.64		&	2.20	&		59/105.1		&		30/51.6 		&		33/67.5 		&		$-0.12$		&		42.0		&		65.0		&		31.4		&		1.17		&	 \ldots \\
64d			&		J132614.038$-$473019.97		&		0.68		&	2.25	&		84/95.4 		&		27/29.1 		&		61/77.9 		&		$-0.43$		&		42.3		&		56.3		&		19.2		&		2.80		&	 \ldots \\
{\bf 64e}		&		J132640.754$-$473450.95		&		1.16		&	2.39	&		25/26.1 		&		2/0.0   		&		28/35.8 		&		\ldots		&		10.7		&		16.8		&		2.7		&		\ldots		&	 \ldots \\
{\bf 64f}		&		J132622.346$-$473319.09		&		1.41		&	2.40	&		21/20.1 		&		16/18.1 		&		7/1.8   		&		1.01		&		8.4		&		11.5		&		7.6		&		\ldots		&	 \ldots \\
71a			&		J132604.190$-$472805.61		&		1.63		&	2.83	&		37/26.9 		&		16/14.8 		&		23/13.5 		&		0.04		&		12.6		&		15.2		&		7.7		&		0.94		&	 \ldots \\
71b			&		J132604.553$-$472741.47		&		0.78		&	2.82	&		118/150.7		&		56/81.6 		&		73/102.5		&		$-0.10$		&		68.5		&		100.2		&		42.5		&		1.29		&	 \ldots \\ \thispagestyle{empty}
71c			&		J132612.758$-$472411.42		&		0.73		&	2.87	&		119/183.1		&		49/75.4 		&		81/142.0		&		$-0.27$		&		79.3		&		114.8		&		47.4		&		0.96		&	 \ldots \\
71d			&		J132617.590$-$472337.26		&		0.99		&	2.78	&		57/64.4 		&		18/18.9 		&		51/67.2 		&		$-0.55$		&		28.1		&		45.0		&		13.4		&		1.84		&	 \ldots \\
71e			&		J132623.124$-$472250.71		&		0.71		&	2.79	&		128/162.1		&		57/72.1 		&		80/112.4		&		$-0.19$		&		71.7		&		99.5		&		43.8		&		1.38		&	 \ldots \\
{\bf 71f}		&		J132610.478$-$472558.74		&		1.53		&	2.64	&		30/23.2 		&		14/13.6 		&		27/22.3 		&		$-0.21$		&		10.6		&		19.3		&		6.7		&		\ldots		&	 \ldots \\
{\bf 71g}		&		J132633.403$-$472159.40		&		1.16		&	2.78	&		37/40.0 		&		14/18.1 		&		24/23.9 		&		$-0.12$		&		18.3		&		22.5		&		11.5		&		\ldots		&	 \ldots \\
{\bf 71h}		&		J132642.667$-$472125.76		&		1.63		&	2.86	&		30/22.9 		&		9/7.7   		&		25/18.3 		&		$-0.37$		&		10.6		&		13.9		&		2.9		&		\ldots		&	 \ldots \\
72a			&		J132648.007$-$472150.68		&		0.86		&	2.68	&		53/67.1 		&		24/35.1 		&		33/41.9 		&		$-0.08$		&		29.8		&		40.9		&		19.2		&		1.66		&	 \ldots \\
72b			&		J132654.521$-$472204.63		&		0.48		&	2.64	&		214/693.1		&		91/320.7		&		139/513.6		&		$-0.20$		&		325.3		&		448.1		&		199.5		&		0.45		&	 \ldots \\
72d			&		J132723.304$-$472447.17		&		0.77		&	2.82	&		73/96.1 		&		18/22.2 		&		65/100.6		&		$-0.66$		&		39.0		&		60.0		&		17.3		&		0.97		&	 \ldots \\
73a			&		J132721.706$-$473206.73		&		0.74		&	2.60	&		69/81.7 		&		47/57.2 		&		24/25.6 		&		0.35		&		34.4		&		45.7		&		26.4		&		0.69		&	 NV369 \\
73b			&		J132720.923$-$473233.65		&		1.19		&	2.65	&		31/28.0 		&		16/16.8 		&		17/12.0 		&		0.14		&		11.5		&		15.4		&		7.1		&		0.42		&	 \ldots \\
73c			&		J132659.256$-$473457.75		&		0.54		&	2.52	&		249/312.5		&		94/113.0 		&		170/245.3		&		$-0.34$		&		132.3		&		184.9		&		73.1		&		1.68		&	 \ldots \\
73d			&		J132647.316$-$473559.18		&		0.97		&	2.79	&		49/66.6 		&		28/44.3 		&		23/25.9 		&		0.23		&		29.0		&		36.6		&		21.8		&		0.64		&	 (V210) \\
{\bf 73e}		&		J132721.458$-$473221.21		&		1.30		&	2.63	&		27/23.3 		&		3/0.0   		&		31/34.5 		&		\ldots		&		10.0		&		16.2		&		1.2		&		\ldots		&	 \ldots \\
74a			&		J132629.038$-$473436.90		&		1.00		&	2.56	&		39/47.6 		&		25/32.2 		&		15/14.0 		&		0.36		&		19.9		&		26.0		&		13.1		&		0.98		&	 \ldots \\
74b			&		J132627.499$-$473455.37		&		0.58		&	2.71	&		263/374.1		&		114/160.1		&		169/273.8		&		$-0.23$		&		167.5		&		235.0		&		93.7		&		1.02		&	 \ldots \\
74c			&		J132617.285$-$473408.12		&		1.42		&	2.85	&		37/35.5 		&		8/5.5   		&		41/51.2 		&		$-0.97$		&		15.6		&		28.6		&		2.4		&		0.97		&	 \ldots \\
74d			&		J132608.314$-$473032.85		&		1.04		&	2.64	&		56/52.2 		&		25/24.3 		&		33/32.5 		&		$-0.13$		&		23.0		&		30.9		&		14.9		&		0.53		&	 V216 \\
{\bf 74g}		&		J132633.031$-$473449.33		&		1.48		&	2.52	&		20/20.8 		&		2/0.0   		&		22/28.1 		&		\ldots		&		9.1		&		12.9		&		1.5		&		\ldots		&	 \ldots \\
81a			&		J132634.793$-$472055.39		&		1.70		&	3.15	&		31/30.6 		&		16/24.3 		&		20/15.7 		&		0.19		&		14.0		&		21.0		&		11.3		&		0.81		&	 \ldots \\ \thispagestyle{empty}
{\bf 81b}		&		J132614.350$-$472241.71		&		1.59		&	3.19	&		33/39.6 		&		9/10.5  		&		33/46.7 		&		$-0.65$		&		16.9		&		30.1		&		8.1		&		\ldots		&	 \ldots \\
82b			&		J132710.181$-$472127.94		&		0.77		&	3.20	&		112/186.5		&		90/203.5		&		23/28.2 		&		0.86		&		82.4		&		111.8		&		79.2		&		0.98		&	 (NV377) \\
82c			&		J132718.026$-$472256.15		&		1.20		&	3.03	&		39/43.3 		&		7/6.0   		&		37/46.2 		&		$-0.89$		&		18.8		&		26.0		&		5.8		&		0.80		&	 \ldots \\
82d			&		J132721.151$-$472323.35		&		0.64		&	3.05	&		199/277.3		&		92/132.9		&		120/188.4		&		$-0.15$		&		118.3		&		167.3		&		72.7		&		0.99		&	 \ldots \\
82e			&		J132728.366$-$472421.82		&		0.80		&	3.19	&		103/151.4		&		37/56.0 		&		75/127.1		&		$-0.36$		&		63.8		&		92.6		&		34.6		&		0.61		&	 \ldots \\
82f			&		J132729.292$-$472553.77		&		0.69		&	2.97	&		140/178.8		&		53/68.3 		&		95/138.5		&		$-0.31$		&		75.7		&		104.7		&		35.9		&		0.13		&	 \ldots \\
82g			&		J132730.629$-$472654.11		&		1.00		&	2.93	&		54/55.7 		&		13/10.1 		&		44/54.3 		&		$-0.73$		&		23.4		&		32.0		&		7.2		&		0.70		&	 \ldots \\
{\bf 82h}		&		J132719.838$-$472250.40		&		1.10		&	3.14	&		50/61.0 		&		5/4.4   		&		57/84.1 		&		$-1.28$		&		26.3		&		43.6		&		6.3		&		\ldots		&	 \ldots \\
83a			&		J132727.410$-$473132.16		&		0.60		&	2.84	&		214/262.2		&		106/129.4		&		128/175.7		&		$-0.13$		&		112.3		&		159.5		&		73.0		&		0.49		&	 \ldots \\
83b			&		J132720.126$-$473336.01		&		1.18		&	2.85	&		38/38.3 		&		12/11.5 		&		31/35.7 		&		$-0.49$		&		16.0		&		23.7		&		9.3		&		0.82		&	 \ldots \\
83c			&		J132722.834$-$473356.36		&		1.03		&	3.07	&		55/69.5 		&		21/25.3 		&		41/57.8 		&		$-0.36$		&		28.8		&		41.7		&		18.0		&		0.57		&	 \ldots \\ \thispagestyle{empty}
83d			&		J132718.624$-$473403.83		&		0.98		&	2.90	&		52/63.8 		&		10/10.0 		&		54/77.7 		&		$-0.89$		&		27.0		&		43.1		&		10.4		&		0.97		&	 \ldots \\
83e			&		J132712.857$-$473456.86		&		0.53		&	2.92	&		534/691.4		&		230/286.9		&		340/509.1		&		$-0.25$		&		293.0		&		415.3		&		175.3		&		0.60		&	 \ldots \\
{\bf 83f}		&		J132714.906$-$473531.05		&		0.95		&	3.18	&		80/106.7		&		13/14.6 		&		99/163.2		&		$-1.05$		&		43.4		&		86.9		&		7.0		&		\ldots		&	 \ldots \\
84a			&		J132639.202$-$473631.34		&		0.77		&	3.05	&		135/196.4		&		51/86.0  		&		97/153.9		&		$-0.25$		&		88.6		&		128.6		&		54.6		&		0.72		&	 \ldots \\
84b			&		J132638.191$-$473636.39		&		0.96		&	3.09	&		78/107.0		&		29/48.5 		&		54/76.5 		&		$-0.20$		&		50.1		&		67.1		&		27.9		&		1.34		&	 \ldots \\
84c			&		J132613.620$-$473441.39		&		0.67		&	3.17	&		253/483.4		&		107/203.9		&		162/361.6		&		$-0.25$		&		212.8		&		309.4		&		128.8		&		1.06		&	 \ldots \\
84d			&		J132611.527$-$473402.82		&		0.69		&	3.10	&		240/375.9		&		169/269.1		&		78/132.0  		&		0.31		&		164.2		&		227.2		&		129.5		&		0.51		&	 (V167) \\
84f			&		J132600.881$-$472910.33		&		1.36		&	3.04	&		44/48.9 		&		20/24.8 		&		28/35.8 		&		$-0.16$		&		21.9		&		31.3		&		14.4		&		0.73		&	 \ldots \\
{\bf 84g}		&		J132623.947$-$473611.72		&		0.74		&	3.25	&		198/345.9		&		67/127.5		&		155/307.8		&		$-0.38$		&		153.1		&		233.6		&		89.7		&		\ldots		&	 \ldots \\ \thispagestyle{empty}
91a			&		J132618.708$-$472109.75		&		1.15		&	3.49	&		60/109.5		&		32/71.5 		&		28/47.5 		&		0.18		&		48.6		&		64.8		&		41.2		&		1.46		&	 NV379 \\
{\bf 91c}		&		J132606.754$-$472238.04		&		1.74		&	3.56	&		33/52.1 		&		13/26.8 		&		21/29.5 		&		$-0.04$		&		23.6		&		30.6		&		19.7		&		\ldots		&	 \ldots \\
{\bf 92d}		&		J132732.306$-$472510.43		&		1.29		&	3.26	&		29/49.0 		&		7/9.7   		&		25/46.0 		&		$-0.68$		&		20.2		&		27.2		&		6.7		&		\ldots		&	 \ldots \\
{\bf 92e}		&		J132737.622$-$472633.12		&		1.08		&	3.41	&		60/88.0 		&		22/36.3 		&		47/76.9 		&		$-0.33$		&		37.4		&		58.2		&		22.1		&		\ldots		&	 \ldots \\
93b			&		J132734.596$-$473237.58		&		0.75		&	3.44	&		177/306.3		&		68/133.7		&		124/240.5		&		$-0.25$		&		131.0		&		194.8		&		79.1		&		1.66		&	 \ldots \\
93c			&		J132700.310$-$473714.88		&		0.97		&	3.39	&		59/144.4		&		18/51.3 		&		44/117.2		&		$-0.36$		&		65.6		&		89.3		&		43.2		&		0.90		&	 \ldots \\
{\bf 93d}		&		J132734.006$-$473147.43		&		1.07		&	3.27	&		59/79.7 		&		24/36.7 		&		39/53.9 		&		$-0.17$		&		34.1		&		47.1		&		23.2		&		\ldots		&	 \ldots \\
{\bf 93e}		&		J132725.042$-$473501.23		&		1.47		&	3.46	&		39/53.6 		&		14/19.3 		&		27/37.7 		&		$-0.29$		&		20.7		&		29.4		&		12.3		&		\ldots		&	 \ldots \\
{\bf 93f}		&		J132714.426$-$473559.83		&		1.27		&	3.31	&		47/62.7 		&		13/16.5 		&		43/67.7 		&		$-0.61$		&		25.1		&		42.9		&		18.6		&		\ldots		&	 \ldots \\
{\bf 93g}		&		J132709.665$-$473646.21		&		1.86		&	3.43	&		31/35.6 		&		9/9.7   		&		30/42.8 		&		$-0.65$		&		16.0		&		27.1		&		9.9		&		\ldots		&	 \ldots \\
{\bf 93h}		&		J132657.610$-$473703.11		&		1.46		&	3.28	&		35/48.5 		&		15/27.6 		&		27/36.0 		&		$-0.12$		&		20.1		&		33.6		&		11.7		&		\ldots		&	 \ldots \\
94a			&		J132601.579$-$473305.69		&		0.55		&	3.42	&		1018/2329.2		&		272/688.8		&		820/2167.1		&		$-0.50$		&		1006.0		&		1542.4		&		532.7		&		0.68		&	 CH star \\
94b			&		J132557.374$-$473248.30		&		0.97		&	3.62	&		49/246.5		&		20/131.9		&		34/182.2		&		$-0.14$		&		107.6		&		155.9		&		58.0		&		0.41		&	 \ldots \\
{\bf 94c}		&		J132636.670$-$473729.45		&		1.01		&	3.45	&		47/155.9		&		5/16.8  		&		50/182.6		&		$-1.04$		&		66.5		&		105.0		&		17.5		&		\ldots		&	 \ldots \\
{\bf 94d}		&		J132613.992$-$473541.50		&		1.54		&	3.45	&		40/61.1 		&		6/7.6   		&		48/89.2 		&		$-1.07$		&		26.8		&		50.7		&		10.5		&		\ldots		&	 \ldots \\
{\bf 94e}		&		J132611.326$-$473559.04		&		1.15		&	3.65	&		90/153.6		&		32/62.5 		&		67/121.4		&		$-0.29$		&		66.8		&		99.0		&		43.8		&		\ldots		&	 \ldots \\
{\bf 94f}		&		J132610.774$-$473505.38		&		0.90		&	3.42	&		118/229.5		&		14/22.3 		&		126/293.3		&		$-1.12$		&		93.7		&		163.6		&		21.6		&		\ldots		&	 \ldots \\
{\bf 94g}		&		J132559.299$-$473349.48		&		1.64		&	3.69	&		47/72.9 		&		14/25.9 		&		41/71.8 		&		$-0.44$		&		31.6		&		53.1		&		19.2		&		\ldots		&	 \ldots \\
{\bf 103a}		&		J132733.187$-$473432.51		&		1.12		&	3.74	&		66/137.2		&		28/68.0 		&		42/88.9 		&		$-0.12$		&		55.4		&		82.4		&		39.5		&		\ldots		&	 \ldots \\
{\bf 103b}		&		J132730.869$-$473429.09		&		1.64		&	3.61	&		35/50.5 		&		11/17.4 		&		26/35.3 		&		$-0.31$		&		19.3		&		26.8		&		10.4		&		\ldots		&	 \ldots \\ \thispagestyle{empty}
{\bf 104b}		&		J132611.489$-$473708.07		&		0.79		&	3.99	&		148/854.1		&		59/491.0		&		100/609.2		&		$-0.09$		&		401.2		&		582.4		&		259.2		&		\ldots		&	 \ldots \\
{\bf 104c}		&		J132604.152$-$473513.36		&		1.87		&	3.76	&		43/60.5 		&		17/31.1 		&		35/51.6 		&		$-0.22$		&		25.4		&		44.6		&		20.5		&		\ldots		&	 \ldots \\
{\bf 113a}		&		J132729.333$-$473634.52		&		1.31		&	4.09	&		74/147.8		&		15/29.7 		&		67/151.7		&		$-0.71$		&		60.0		&		93.9		&		26.8		&		\ldots		&	 \ldots \\
{\bf 113b}		&		J132725.066$-$473655.24		&		2.46		&	4.01	&		30/40.0 		&		10/15.7 		&		23/26.3 		&		$-0.22$		&		15.5		&		21.7		&		14.1		&		\ldots		&	 \ldots \\
\enddata
\tablecomments{$^a$Source IDs as assigned in this work (bold face) and in \citet{Haggard2009}; $^b$Source positions in the format JHHMMSS.sss$-$DDMMSS.ss; $^c$95\% confidence error circle radius calculated using Eq.(5) of \citet{Hong2005}; $^d$Offset from \citet{Anderson2010} cluster center (R.A. $=$ 13:26:47.24, Dec. $=$ $-$47:28:46.45) in units of the core radius (\rc\ $=$ 155 arcsec); $^e$ Unabsorbed fluxes determined assuming a power-law spectrum with photon index $\Gamma = 1.4$; $^f$Optical identifications from \citet{Haggard2009}, \citet{Cool2013} and this work; ``NV'' and ``V'' IDs are variable stars from \citet{Kaluzny2004}; parentheses indicate stars are non-members according to \citet{Bellini2009}; $^g$ Fluxes determined using a spectrum appropriate for this object shows no change from 2000 to 2012 \citep{Heinke2014}.}\label{table:master}
\end{deluxetable}